\documentclass[
    ,final            % use final for the camera ready runs
  ]
  {aipproc}

\layoutstyle{8x11single}

\newcommand{\beqn}{\begin{eqnarray}}
\newcommand{\eeqn}{\end{eqnarray}}
\def\beq{\begin{eqnarray}}
\def\eeq{\end{eqnarray}}
\def\CR{\nonumber \\ }

\def \neut {\tilde{\chi}_1^0}
\def \cha {\tilde{\chi}_1^\pm}
\def \mchi {m_{\neut} }
\def \stau {\tilde{\tau}_1}
\def \stop {\tilde{t}_1}

\usepackage{amssymb}

\begin{document}

\title{Dark Matter and EWSB Naturalness in Unified SUSY Models}


\author{Pearl Sandick}{
  address={University of Utah, Dept. of Physics, Salt Lake City, UT 84112, USA}
}

\begin{abstract}

The relationship between the degree of fine-tuning in Electroweak Symmetry Breaking (EWSB) and the discoverability
of dark matter in current and next generation
direct detection experiments is investigated in the context of two
unified Supersymmetry scenarios: the constrained
Minimal Supersymmetric Standard Model (CMSSM)
and models with non-universal Higgs masses
(NUHM).  Attention is drawn to the mechanism(s) by which the relic abundance of neutralino dark matter is suppressed to cosmologically viable values.
After a summary of Amsel, Freese, and Sandick (2011), results are updated to reflect current constraints, including the discovery of a new particle consistent with a Standard Model-like
Higgs boson.  We find that a Higgs mass of $\sim125$ GeV excludes the least fine-tuned CMSSM points in our parameter space and that remaining viable models may be difficult to probe with next generation direct dark matter searches.  Relatively low fine-tuning and good direct detection prospects are still possible in NUHM scenarios.

\end{abstract}

\maketitle

%%%%%%%%%%%%%%%%%%%%%%%%%%%%%%%%%%%%%%%%%%%%
%% MAINMATTER
%%%%%%%%%%%%%%%%%%%%%%%%%%%%%%%%%%%%%%%%%%%%

\section{Introduction}

Supersymmetry is a well-studied and elegant extension of the Standard Model of particle physics.  If supersymmetry is broken near the weak scale, not only is it possible to achieve gauge coupling unification, but the Hierarchy Problem is also addressed through cancellations of the quadratically divergent corrections to the Higgs mass squared.  Furthermore, supersymmetric theories provide a host of natural particle candidates for dark matter.  Specifically, we are interested in supersymmetric models in which the lightest neutralino is the lightest supersymmetric particle (LSP), and constitutes some or all of the observed dark matter in the universe.  In many cases, neutralino dark matter is predicted to scatter on nuclei with a cross section large enough to be observed by current or next generation direct detection experiments.

Despite these virtues, supersymmetric theories often suffer from fine-tuning issues related to the fact that the Z mass can be calculated from supersymmetric parameters that may take values quite far from the weak scale.  Here we address the question of what can be learned about fine-tuning from direct dark matter searches.

I begin by summarizing the results of Amsel, Freese, and Sandick (2011)~\cite{AFS}, in which ease of discoverability of neutralino dark matter in direct detection experiments, fine-tuning of electroweak symmetry breaking (EWSB), and the relationship between the two are explored for two supersymmetric scenarios in which there are large degrees of universality at the supersymmetric grand unification (GUT) scale.  The conclusions reached are in agreement with those of Ref.~\cite{Perelstein:2011tg}, which investigated the MSSM with relevant parameters specified at the weak scale and with the assumption that neutralinos constitute all of the dark matter in the Universe.  After presenting the results of~\cite{AFS}, I go on to explore the sensitivity of those conclusions to the mass of the Standard Model-like Higgs boson~\cite{higgsCMS,higgsATLAS} and the recently improved measurement of the branching ratio, $BR(B_s \rightarrow \mu^+ \mu^-)$~\cite{LHCbmm}.

The two scenarios studied are special cases of the Minimal Supersymmetric Standard Model (MSSM); namely, the constrained MSSM (CMSSM), and models with non-universal Higgs masses, (NUHM).  All CMSSM models can be described by four parameters and a sign: a universal mass for all gauginos, $M_{1/2}$, a universal mass for all scalars, $M_0$, a universal trilinear coupling, $A_0$, the ratio of the Higgs vacuum expectation values, $\tan \beta$, and the sign of the Higgs mass parameter, $\mu$.  In NUHM scenarios, additional freedom in the Higgs sector is introduced by allowing the supersymmetry-breaking contributions to the effective masses of the up- and down-type Higgs scalars, $m_{H_u}$ and $m_{H_d}$, respectively, to differ from the universal mass taken by the supersymmetric scalar particles at the GUT scale, $M_0$.  

After introducing the methodology and parameter space, the relationship between EWSB fine-tuning and discoverability of dark matter via neutralino-nucleon elastic scattering is discussed, with emphasis on the specific mechanisms by which the relic abundance of neutralino dark matter is reduced to cosmologically viable values in the CMSSM and the NUHM.  Finally, the impact of updated constraints on the conclusions of~\cite{AFS} is addressed.

%%%%%%%%%%%%%%%%%%%%%%%%%%%%%%%%%%%%%%%
\section{The Parameter Space}

In Ref.~\cite{AFS}, we explore the parameter space of the CMSSM and NUHM by scanning over the relevant input parameters, applying constraints, and identifying trends in viable models. In both the CMSSM and the NUHM, we assume $\mu >0$ and scan the ranges $1 < \tan\beta < 60$ and $-12$ TeV $< A_{0} < 12$ TeV. In the CMSSM, we scan $0 < M_{0} < 4$ TeV and $0 < M_{1/2} < 2$ TeV while in NUHM space we take $0 < M_{0} < 3$ TeV, $0 < M_{1/2} < 2$ TeV\footnote{For NUHM models, our scan is more dense for $M_{1/2} < 1$ TeV.  The difference in density of points does not affect our conclusions.}, and the GUT-scale Higgs scalar mass parameters $-3$ TeV $< M_{H_{u,d}}(M_{GUT}) < 3$ TeV.   Since gaugino universality is assumed in both cases, the electroweak scale mass relations of $M_{1} : M_{2} : M_{3} \approx 1 : 2 : 6$ hold for both the CMSSM and the NUHM.

We begin by imposing a lower limit on the mass of the light CP-even Higgs boson, $m_{h} > 114$ GeV~\cite{LEPhiggs}.  Accelerator bounds on SUSY parameters are enforced, including $m_{\cha} > 104$ GeV~\cite{Abbiendi:2003sc} and, following~\cite{Feldman:2009zc}, $m_{\tilde{t}_1, \tilde{\tau}_1} > 100$ GeV.  We allow a $3\sigma$ range for the $BR(b\rightarrow s\gamma)$ as recommended by the Heavy Flavor Averaging Group~\cite{hfag}; accounting for the improved Standard Model calculation~\cite{bsgSM}, we take $2.77\times 10^{-4}< BR(b\rightarrow s\gamma) <4.27\times 10^{-4}$.  We also demand demand $-11.4\times 10^{-10}< \delta(g_{\mu}-2)<9.4\times 10^{-9}$~\cite{Djouadi:2006be}.  Finally, we require $BR( B_s \to \mu^{+}\mu^{-})  < 10^{-7}$ as measured by CDF~\cite{CDFbmumu}.  After summarizing the results of~\cite{AFS} as obtained by implementing these constraints, we explore the sensitivity of our conclusions to the Higgs mass and the recently improved limit on $BR( B_s \to \mu^{+}\mu^{-})$.

For all models, we apply the $2\sigma$ upper limit on the relic abundance of neutralino dark matter\footnote{In~\cite{AFS} we also investigate the implications of requiring that the entire dark matter abundance is due to neutralinos.} of $\Omega_{\neut} h^{2} < 0.12$~\cite{WMAP7}.  For bino-like neutralino LSPs, thermal freeze out typically results in an overabundance of dark matter. Regions of parameter space where the abundance falls in or below the observed range often involve one or more specific mechanisms that act to reduce the neutralino abundance.  Examples of these mechanisms include coannihilations of LSPs with other supersymmetric particles and LSP annihilations at a Higgs pole.  In the former case, it's necessary that a viable coannihilation candidate particle be nearly degenerate in mass with the neutralino LSP, while in the latter case, s-channel annihilations through $h$ or $A$ exchange are enhanced when $2m_{\neut} \approx m_{h,A}$.  If the LSP has a significant higgsino admixture, it is possible for the relic density to be consistent with the measured abundance of dark matter even in the absence of coannihilations or a resonance.  We make no a priori assumptions about the composition of the neutralino LSP, and we differentiate model categories by the mass relation obeyed by the relevant particles as listed in Table~\ref{tab:degeneracies}.  We label the model categories by annihilation mechanisms; the named mechanism is usually, but not always, the primary one for producing the correct relic abundance.  In some cases, models may satisfy more than one mass relation, while in other cases, none of the mass relations are satisfied.  If the latter, points that obey the limit on the relic density of dark matter are labeled ``other,'' and typically the LSP is a mixed bino-higgsino state.
 
 \begin{table}[h]
\begin{tabular}{lc p{5mm}c lc}
\hline
  \tablehead{1}{l}{b}{Model Category}
  & \tablehead{1}{c}{b}{Mass Relation} 
   & \tablehead{1}{c}{b}{ } 
  & \tablehead{1}{l}{b}{Model Category}
  & \tablehead{1}{c}{b}{Mass Relation}\\
\hline 
Stop Coannihilation & $m_{\tilde{t}_1} - \mchi  < 0.2 \mchi$ & & Light Higgs Pole & $| \frac{1}{2}m_h - \mchi | < 0.1 \mchi$ \\
Stau Coannihilation & $m_{\tilde{\tau}_1} - \mchi  < 0.2 \mchi$ & & Heavy Higgs Pole & $| \frac{1}{2}m_A - \mchi | < 0.1 \mchi$ \\
Chargino Coannihilation & $m_{\tilde{\chi}_1^\mp} - \mchi  < 0.15 \mchi$ & & & \\
\hline
\end{tabular}
\caption{Annihilation mechanisms as specified by mass relations.  All model points that do not obey one of these mass relations are labeled "other."}
\label{tab:degeneracies}
\end{table}

\begin{figure}
  \includegraphics[width=.5\textwidth, height=65mm]{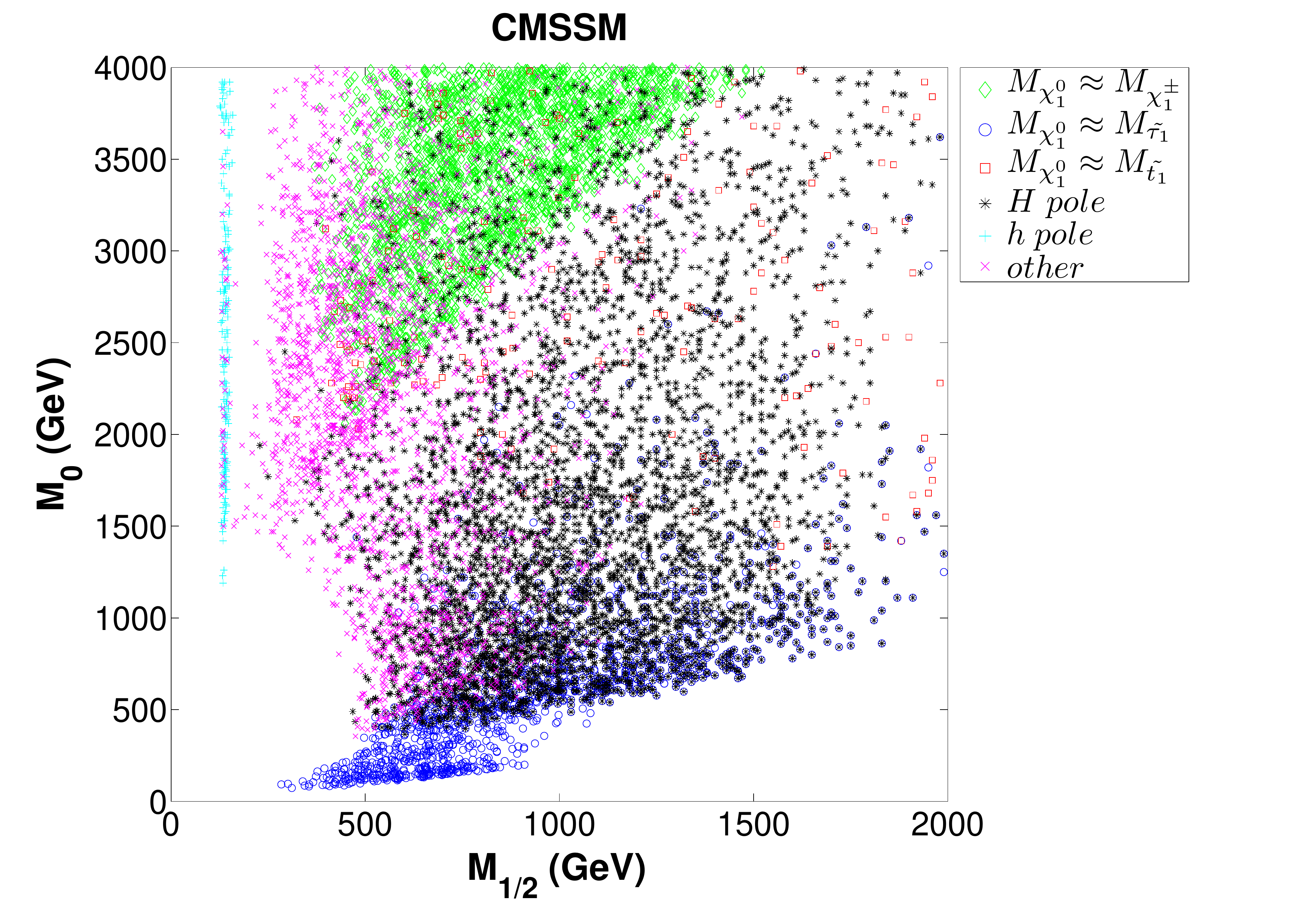}
   \includegraphics[width=.5\textwidth, height=65mm]{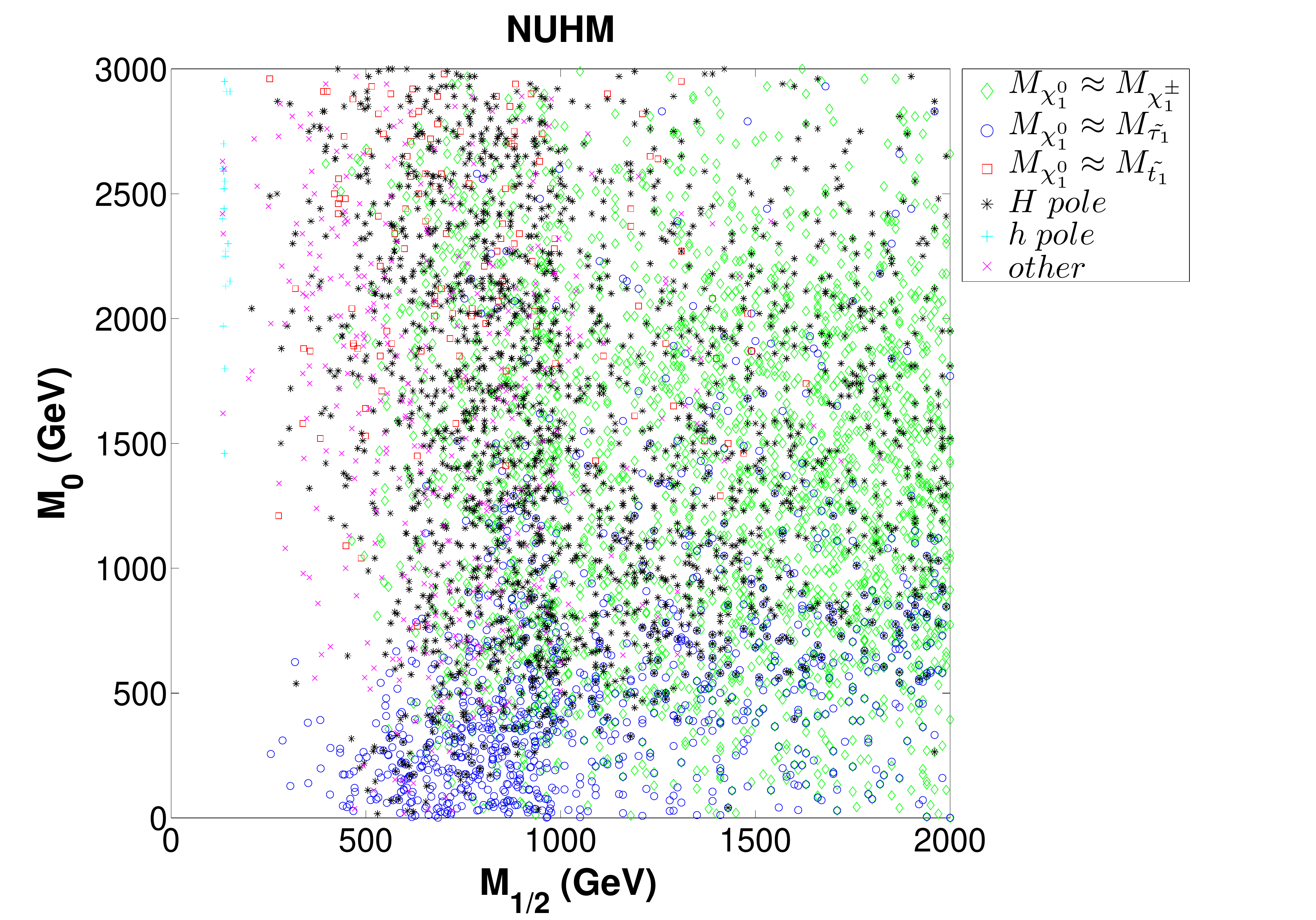}
  \caption{The $(M_{1/2},M_0)$ plane of the CMSSM (left) and the NUHM (right).  
Models are color-coded by mass relation as described in the legend.} \label{fig:m0_mhf}\end{figure}

In Figure~\ref{fig:m0_mhf} we present the $(M_{1/2},M_0)$ plane of the CMSSM (left) and the NUHM (right).  Of the mass relations plotted, some are more localized in the CMSSM plane than in the NUHM plane. For example, the $m_{\neut}\approx m_{\cha}$ points in the CMSSM all occur at large $M_0$ and small $M_{1/2}$ since that is the only region of the CMSSM plane where the neutralino LSP is significantly higgsino-like so that this near-degeneracy is possible. In the NUHM, however, the restriction that $m_{H_u}(M_{GUT})=m_{H_d}(M_{GUT})=M_0$ is relaxed, so there is considerably more freedom in the Higgs sector.  As a result, the neutralino LSP may be higgsino-like in any region of the $(M_{1/2},M_0)$ plane.  Indeed, there are $m_{\neut}\approx m_{\cha}$ points spread throughout the NUHM plane in the right panel of Fig.~\ref{fig:m0_mhf}.

%%%%%%%%%%%%%%%%%%%%%%%%%%%%%%%%%%%%%%%
\section{Fine-tuning}

Despite the successes of the MSSM, fine-tuning of the Z mass is a generic issue for supersymmetric models.
Neglecting loop corrections, the Z mass in the MSSM is given by 
\beq 
\label{zmass}
m_Z^2 =
\frac{|m_{H_d}^2-m_{H_u}^2|}{\sqrt{1-\sin^2 2\beta}}-m_{H_d}^2-m_{H_u}^2-2|\mu|^2,
\eeq
where all parameters are defined at $m_Z$.
Clearly, a cancellation of the terms on the right hand side
is required in order to obtain the measured value of $m_Z$.  However typical values for parameters on the right hand side can be orders of magnitude from the weak scale. 

As noted in~\cite{Ellis:1986yg} and~\cite{BarbieriGiudice}, the degree of fine-tuning may be quantified using log-derivatives.  Here, we follow~\cite{Perelstein:2007nx} and compute the quantity 
\beq 
A(\xi)\,=\,\left| \frac{\partial\log m_Z^2}{\partial\log \xi}\right|,
\label{eq:ftpars} 
\eeq
where $\xi=m_{H_u}^2$, $m_{H_d}^2$, $b$, and $\mu$ are the relevant Lagrangian parameters.  
Then 
\beqn
 A(\mu) &=&
\frac{4\mu^2}{m_Z^2}\,\left(1+\frac{m_A^2+m_Z^2}{m_A^2} \tan^2 2\beta
\right), \CR A(b) &=& \left( 1+\frac{m_A^2}{m_Z^2}\right)\tan^2
2\beta, \CR A(m_{H_u}^2) &=& \left| \frac{1}{2}\cos2\beta
+\frac{m_A^2}{m_Z^2}\cos^2\beta
-\frac{\mu^2}{m_Z^2}\right|\times\left(1-\frac{1}{\cos2\beta}+
\frac{m_A^2+m_Z^2}{m_A^2} \tan^2 2\beta \right), \CR A(m_{H_d}^2) &=&
\left| -\frac{1}{2}\cos2\beta +\frac{m_A^2}{m_Z^2}\sin^2\beta
-\frac{\mu^2}{m_Z^2}\right|\times\left|1+\frac{1}{\cos2\beta}+
\frac{m_A^2+m_Z^2}{m_A^2} \tan^2 2\beta \right|, \CR 
\label{eq:deltapieces}
\eeqn
where it is assumed that $\tan\beta>1$. 
The overall fine-tuning $\Delta$ is
defined as
\beq
\Delta = \sqrt{A(\mu)^2+A(b)^2+A(m_{H_u}^2)^2+A(m_{H_d}^2)^2},
\label{eq:delta}
\eeq
with values of $\Delta$ far above one indicating significant fine-tuning. Quantum corrections further contribute to the fine-tuning, 
e.g.~the one-loop contribution to the $m_{H_u}^2$ parameter from top and stop loops. In this study, we compute the fine-tuning parameter, $\Delta$, accurate to at least one loop, as well as the cross sections for scattering of neutralino dark matter on nuclei, with MicrOMEGAs~\cite{micromegas}, employing the spectrum calculator SUSPECT~\cite{suspect}.

%%%%%%%%%%%%%%%%%%%%%%%%%%%%%%%%%%%%%%%%%
\section{Direct Dark Matter Searches}

For each viable model point, we calculate the spin-independent neutralino-nucleon elastic scattering cross section, $\sigma_{SI}$, to be compared with the limits from direct dark matter searches.  Here, we focus on the XENON-100 limit~\cite{xenon100} and the projected sensitivity of the XENON-1T experiment~\cite{xenon1t}.  For details of the calculation and relevant uncertainties, see~\cite{AFS,Falk:1998xj,Ellis:2008hf}.

\begin{figure}
  \includegraphics[width=65mm,height=63mm]{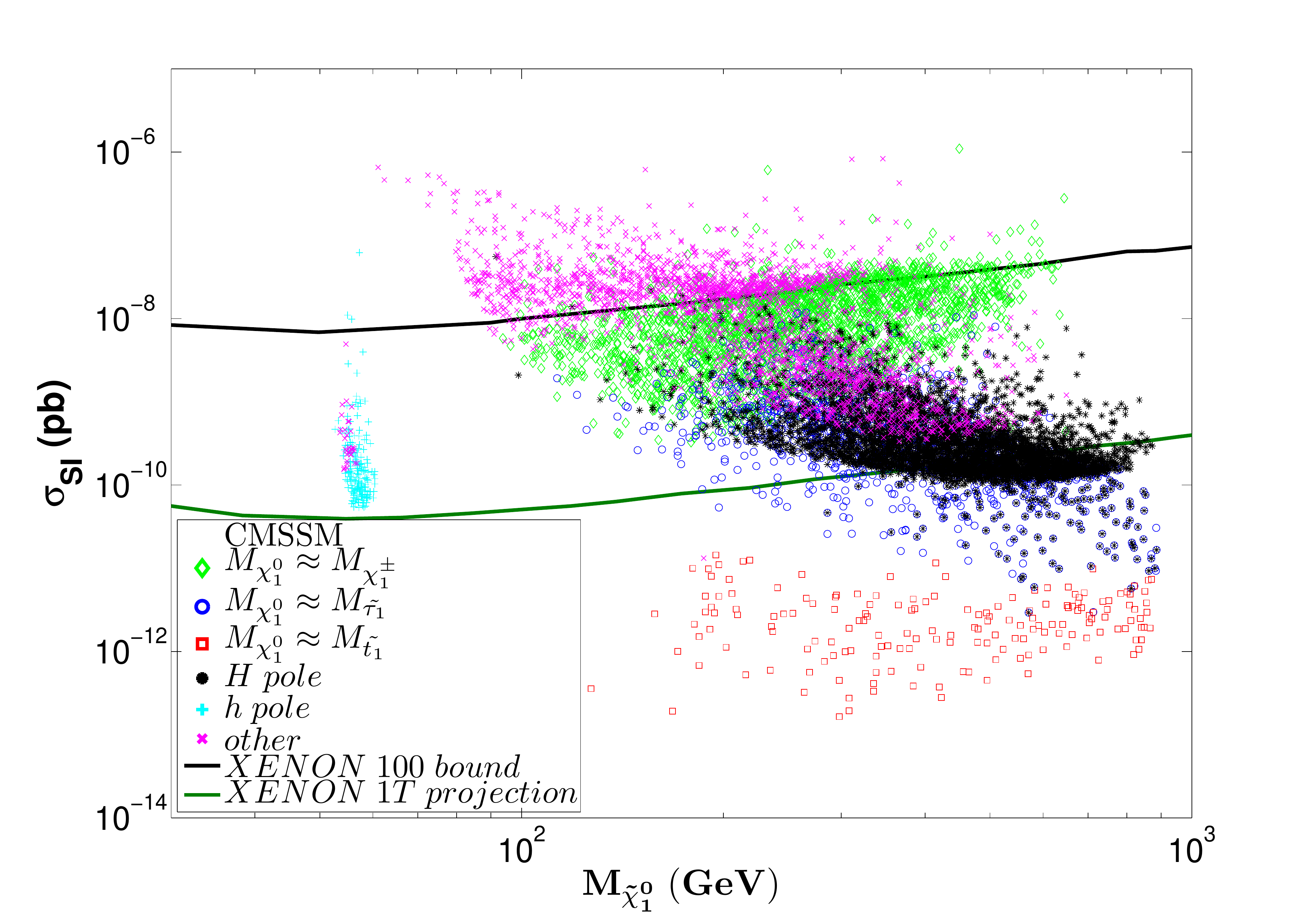}
   \includegraphics[width=65mm, height=63mm]{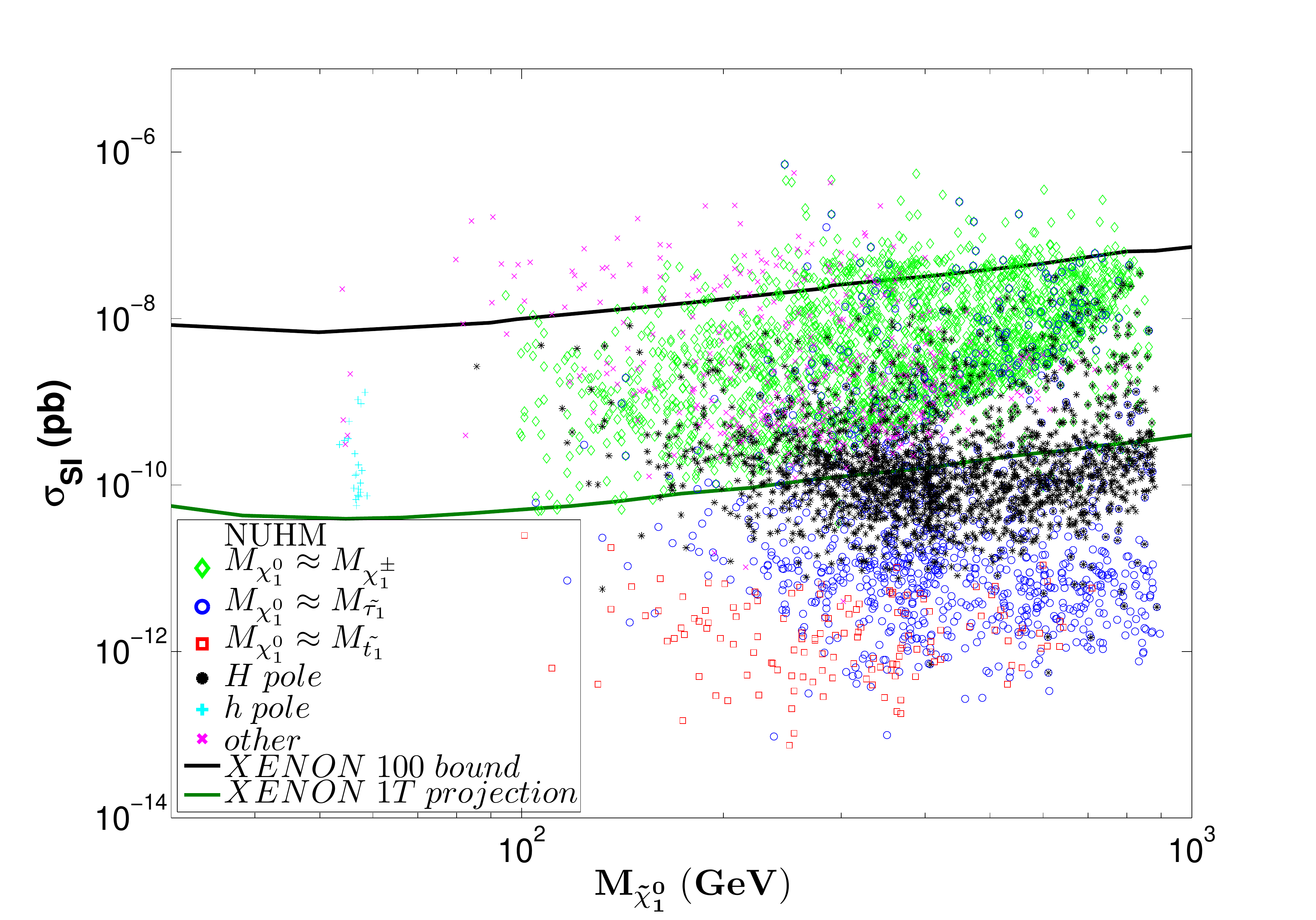}
  \caption{Spin-independent neutralino-nucleon elastic scattering cross section, $\sigma_{SI}$, as a function of neutralino mass $M_{\neut}$ for the CMSSM (left panels) and the NUHM (right panels). Model points are color-coded by mass relation as indicated in the legend. The limit on $\sigma_{SI}$ from XENON-100 and the projected sensitivity of XENON-1T are shown as black and green curves, respectively.
} \label{fig:mchisigsi}
\end{figure}

In Fig.~\ref{fig:mchisigsi} we show $\sigma_{SI}$ as a function of LSP mass, $m_{\neut}$, for model points that pass all constraints, as in~\cite{AFS}. Points are color-coded by mass relation as indicated in the legend. The black (upper) and green (lower) curves in each panel represent the upper limit on $\sigma_{SI}$ from XENON-100 and the projected sensitivity of XENON-1T, respectively. We find that there is significantly more variation in $\sigma_{SI}$ in the NUHM than in the CMSSM, especially for $m_{\neut} \lesssim 150$ GeV or $m_{\neut} \gtrsim 700$ GeV.  This is a straightforward consequence of the additional freedom in the Higgs sector in the NUHM for two reasons: First, since $\mu$ is fixed by the electroweak vacuum conditions, which are related to the Higgs scalar masses, the LSP can be made Higgsino-like for nearly all choices of $M_{1/2}$ and $M_0$ in the NUHM.  Furthermore, in the NUHM it is possible to maintain nearly the measured value of the relic density of neutralinos even if they are largely higgsino-like. In the CMSSM, higgsino-like LSPs annihilate efficiently, resulting in very small $\Omega_{\neut}$, and would therefore be difficult to observe in a direct detection experiment. In the NUHM, however, the varied higgsino content leads to a much larger range of effective scattering cross sections.  Second, since the mass of the heavy CP-even Higgs boson, H, is not constrained by the choice of $M_0$ in the NUHM as it is in the CMSSM, a larger range of $m_H$ is possible. Since $\sigma_{SI} \propto 1/m_H^4$ for scattering via Higgs exchange, a larger range of $\sigma_{SI}$ is therefore possible.  Higgs masses are bounded from below by collider constraints, so the main effect is that since $m_H$ can be much larger in the NUHM than in the CMSSM, lower scattering cross sections are possible. These findings are consistent with those presented in~\cite{nuhm1}.

\begin{figure}
\begin{tabular}{cc}
\includegraphics[width=65mm, height=63mm]{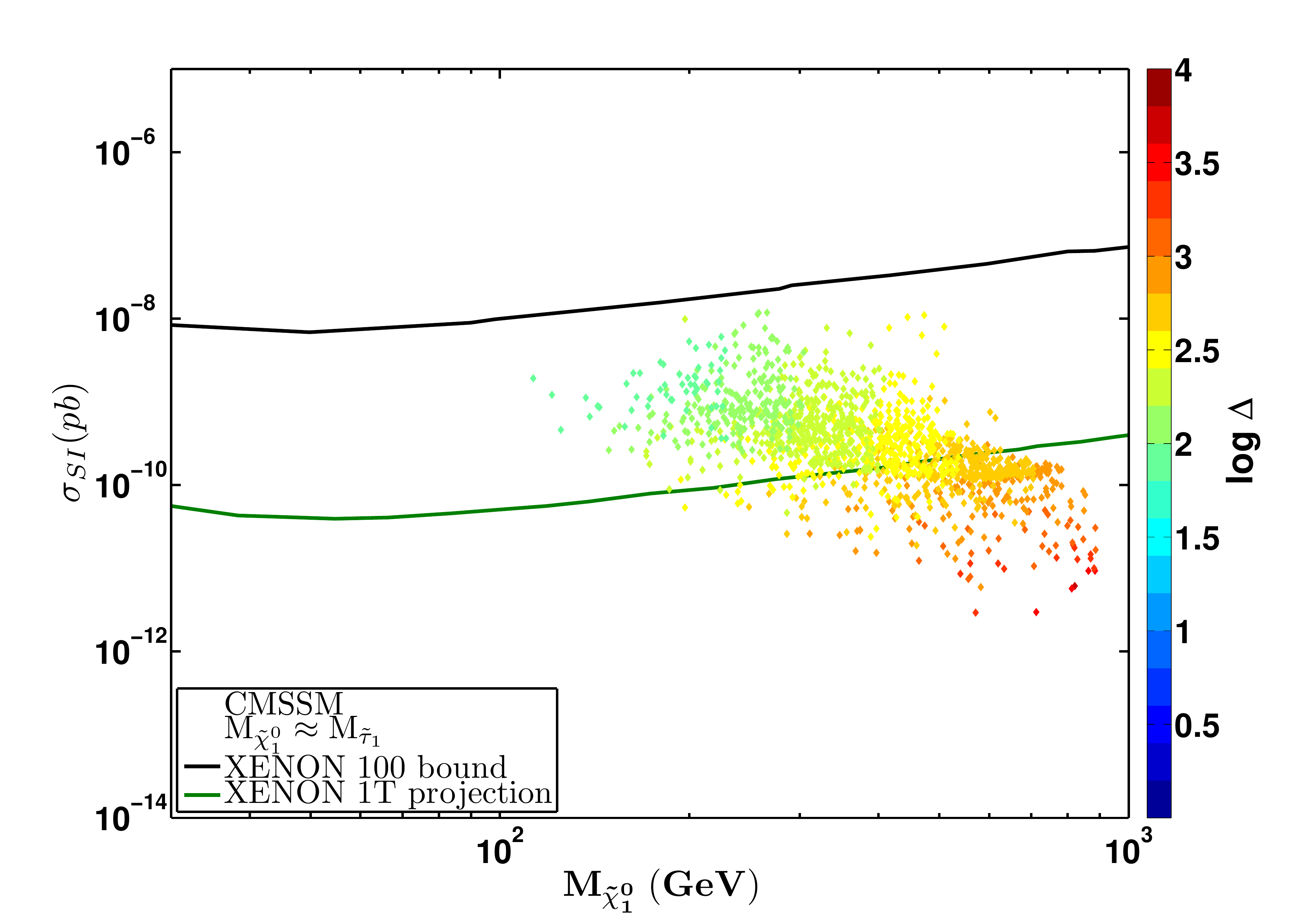} &
\includegraphics[width=65mm, height=63mm]{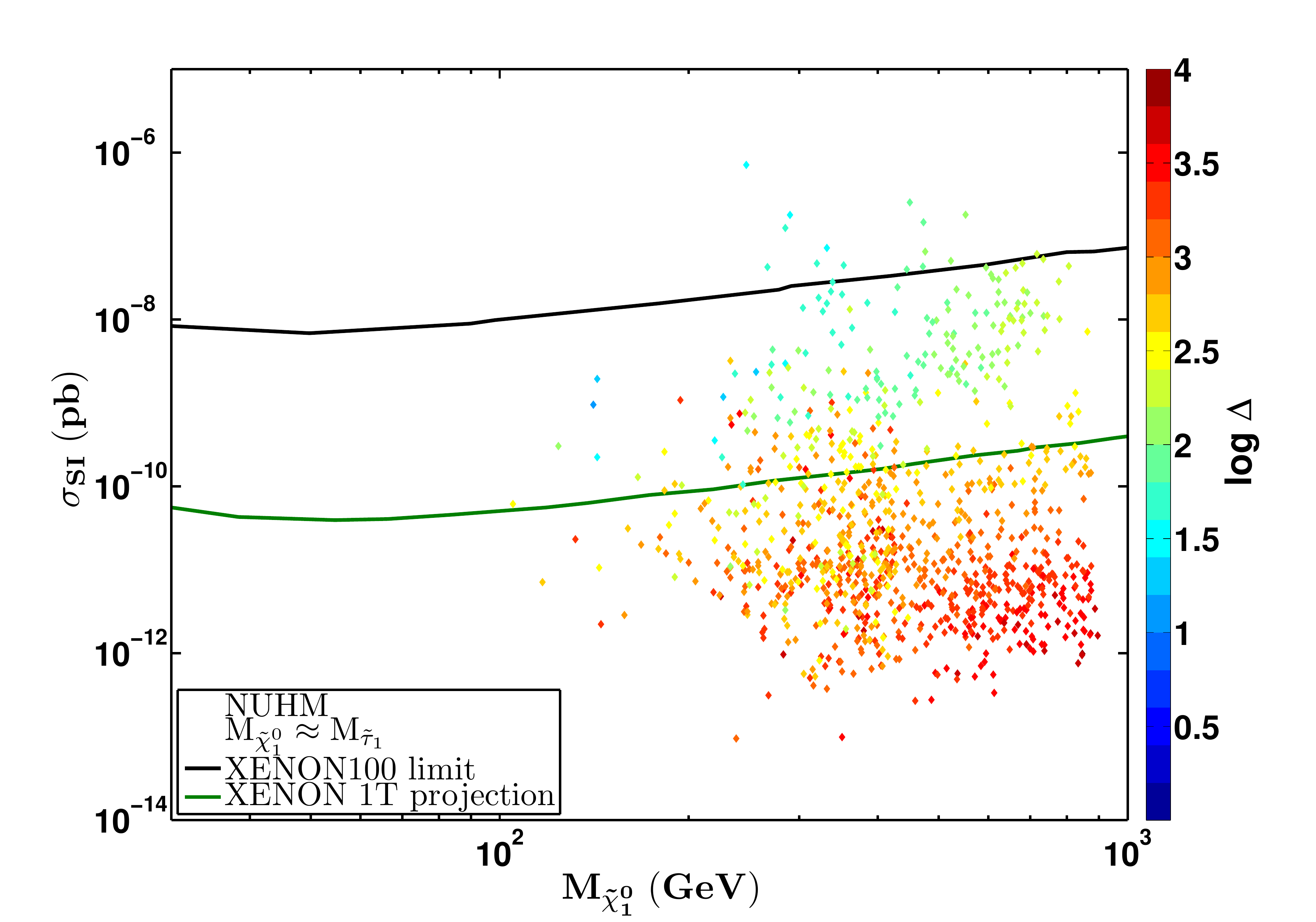} \\
\includegraphics[width=65mm, height=63mm]{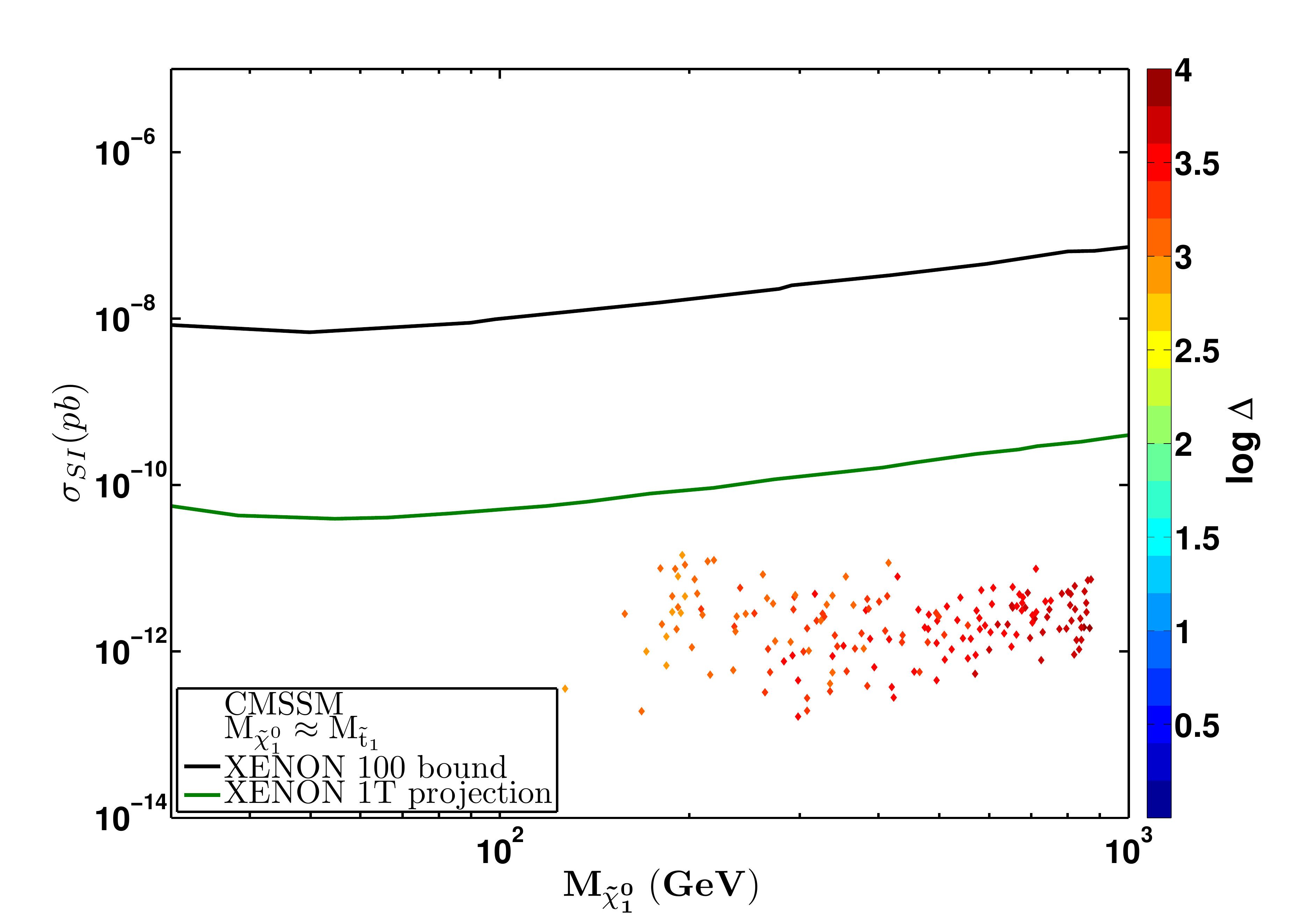} &
\includegraphics[width=65mm, height=63mm]{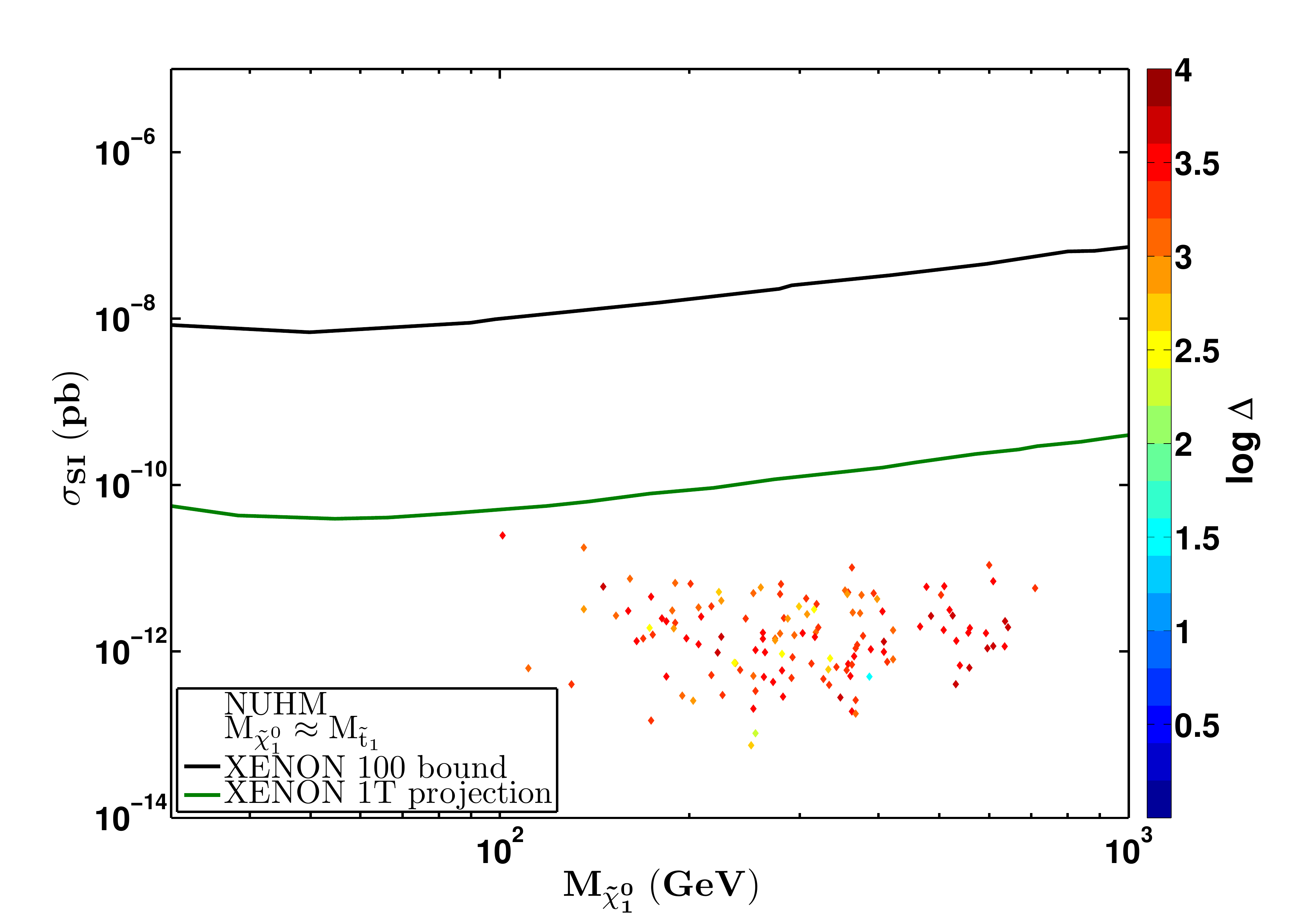}
\end{tabular}
\caption{Spin independent elastic scattering cross section, $\sigma_{SI}$, as a function of LSP mass, $M_{\neut}$, for the CMSSM (left panels) and the NUHM (right panels).  Models are divided according to the mass relations as discussed in the text.  
}
\label{fig:coann}
\end{figure}

\begin{figure}
\begin{tabular}{cc}
\includegraphics[width=65mm, height=63mm]{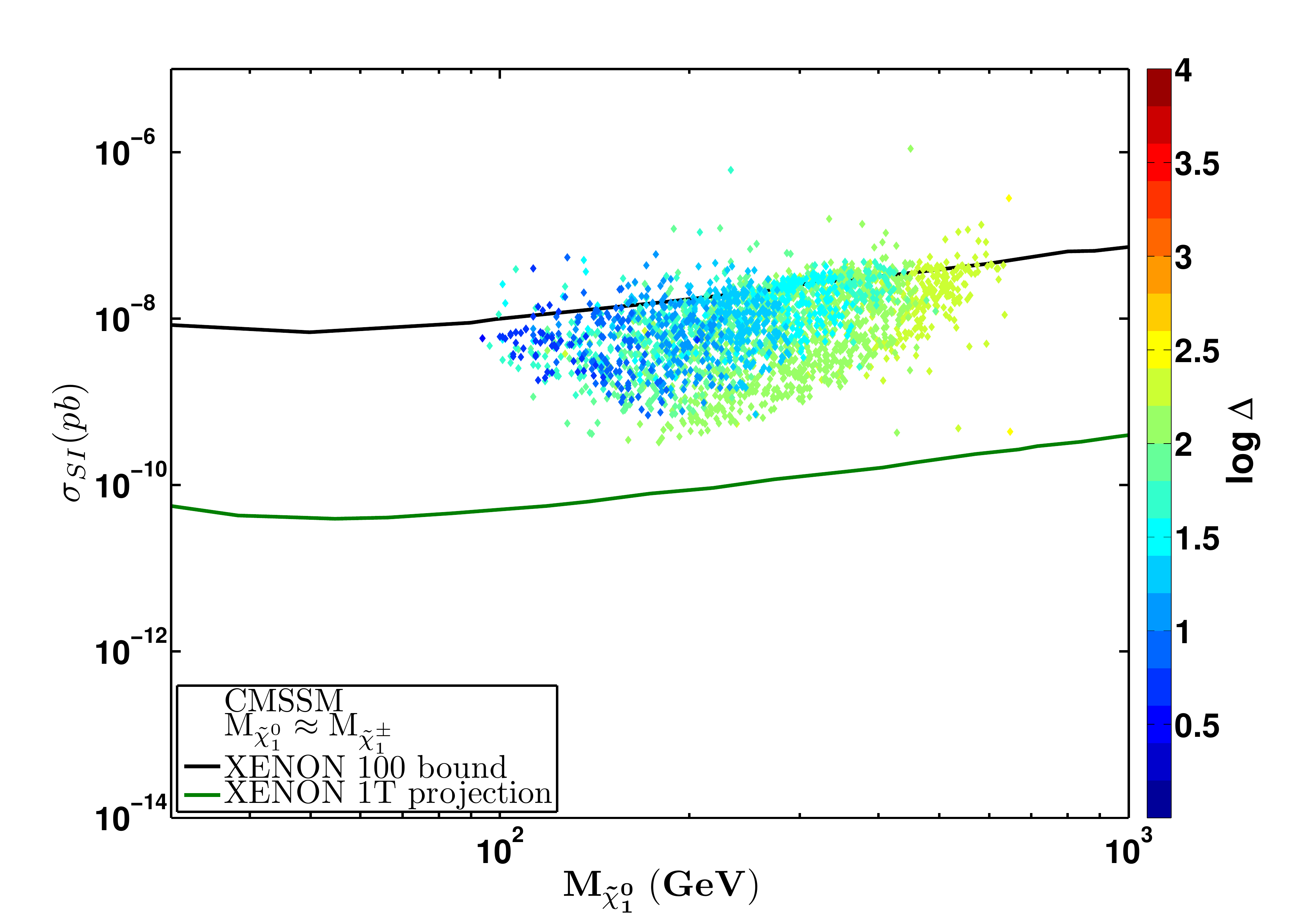} &
\includegraphics[width=65mm, height=63mm]{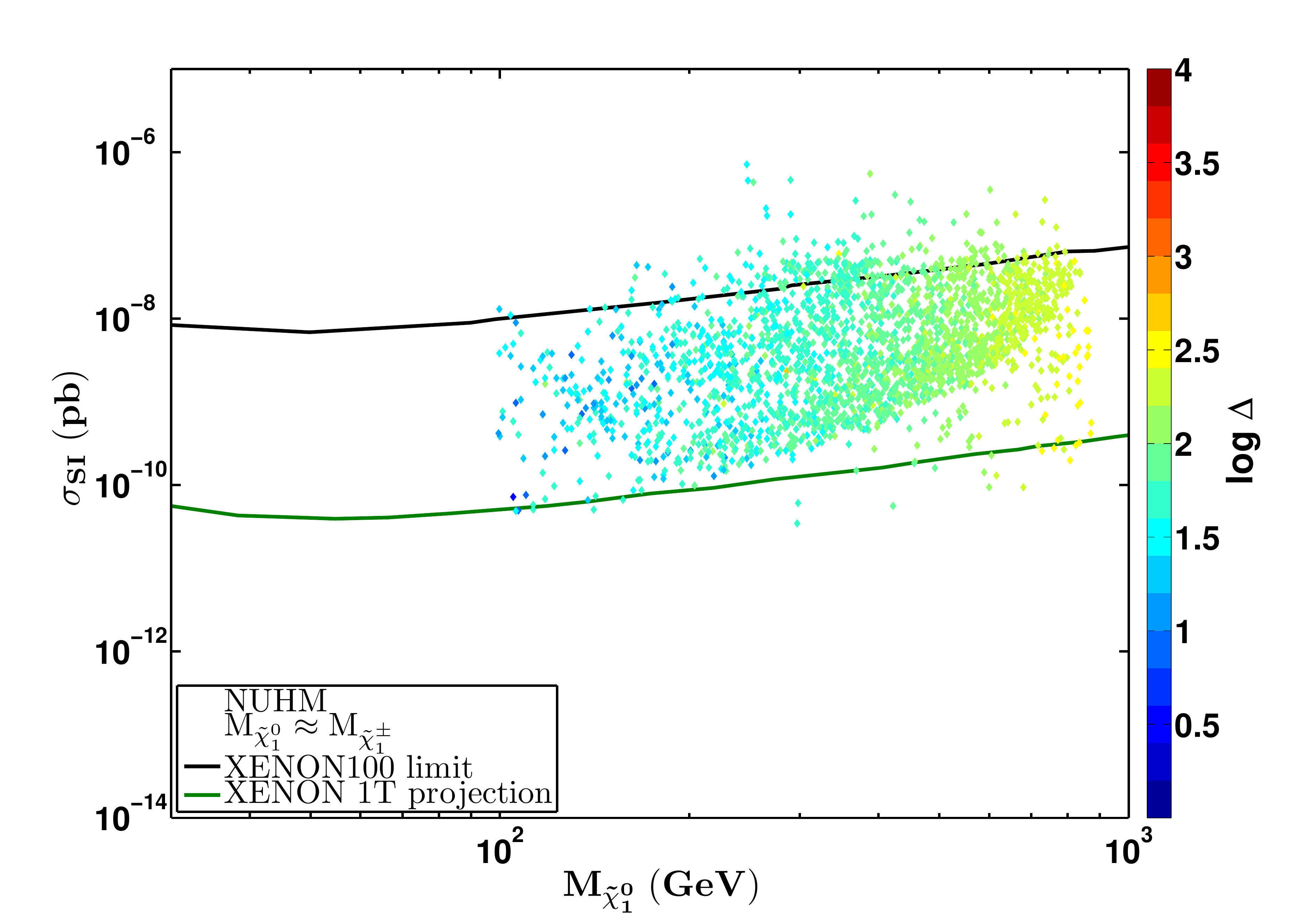} \\
\includegraphics[width=65mm, height=63mm]{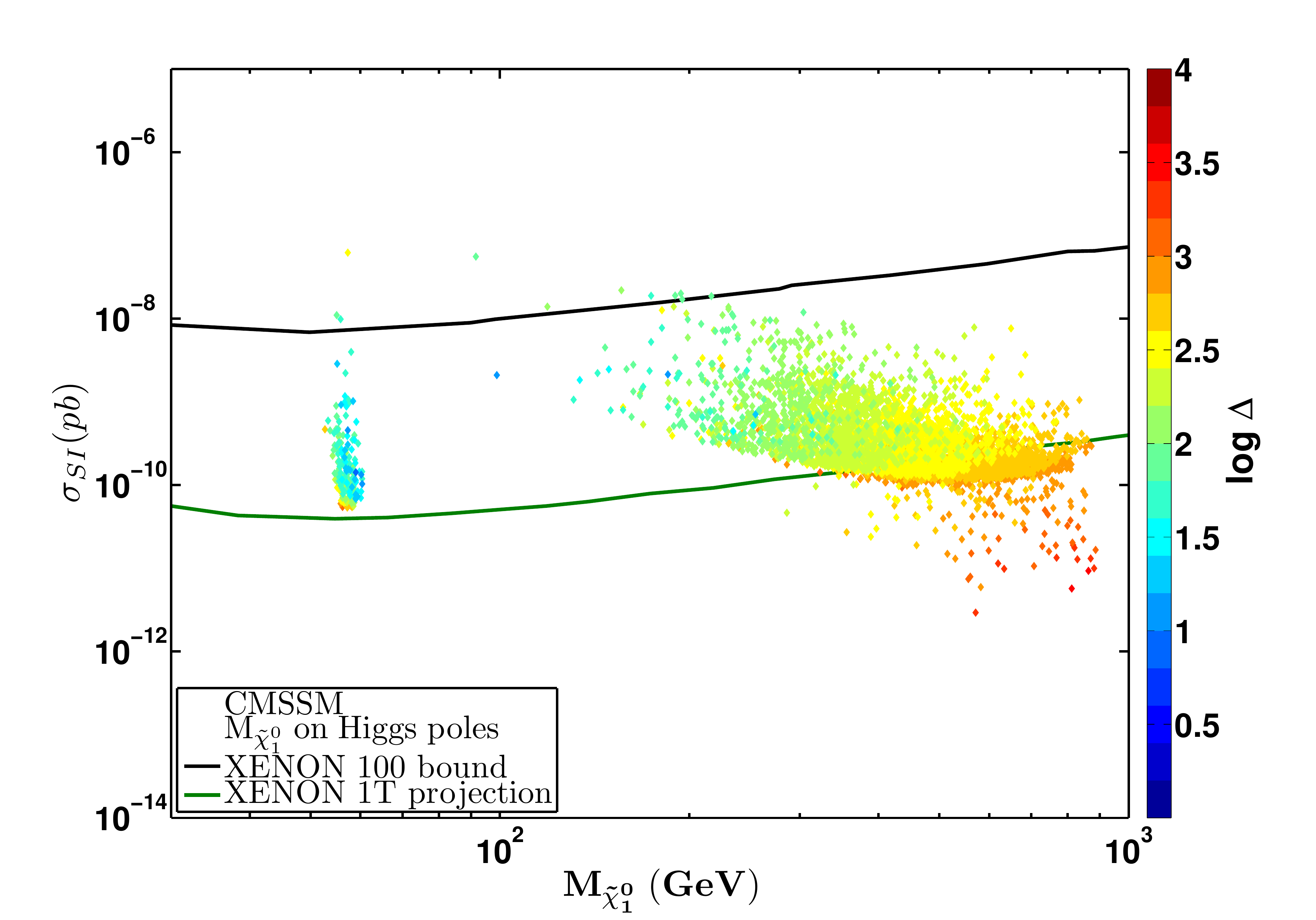} &
\includegraphics[width=65mm, height=63mm]{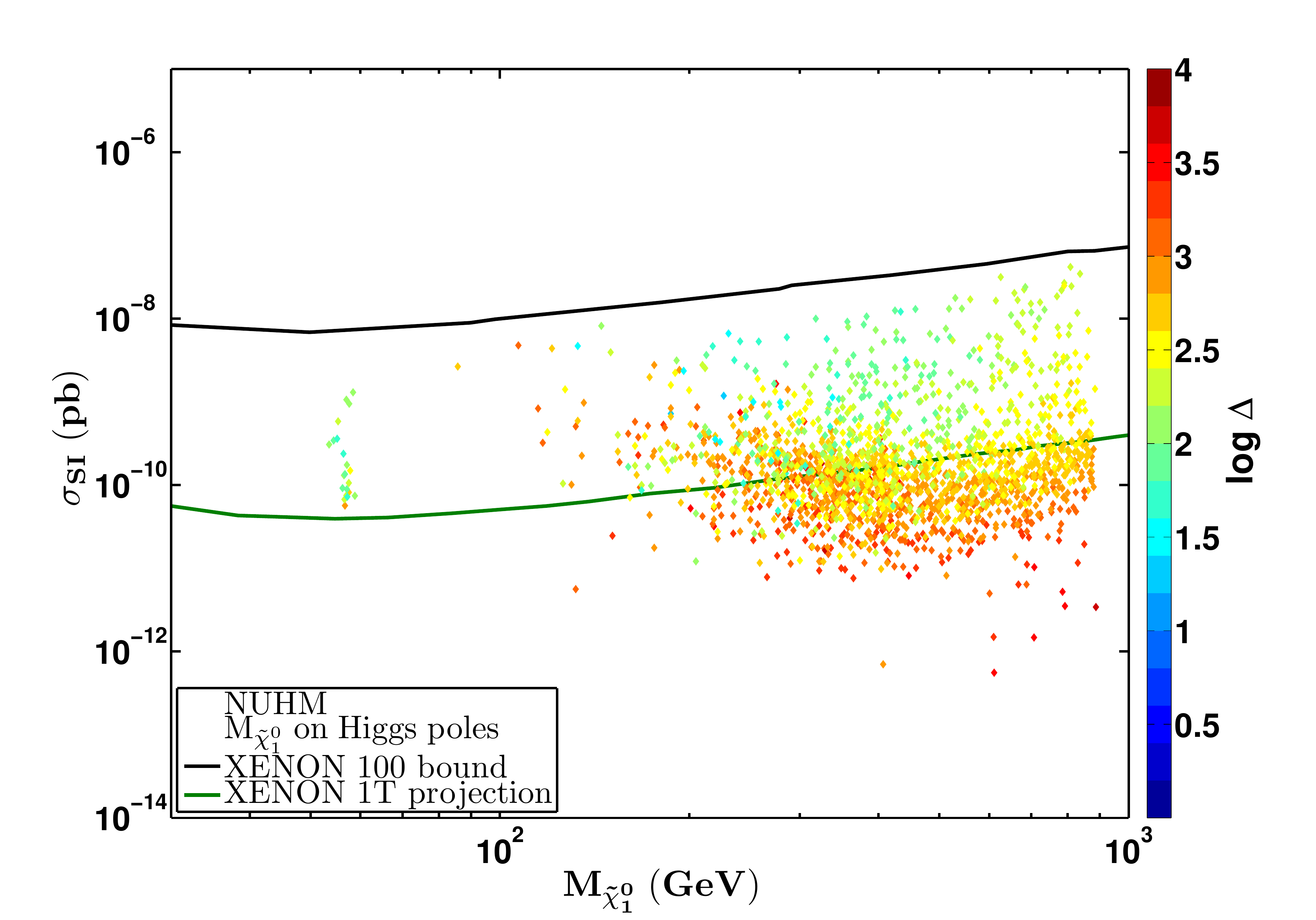}
\end{tabular}
\caption{Spin independent elastic scattering cross section, $\sigma_{SI}$, as a function of LSP mass, $M_{\neut}$, for the CMSSM (left panels) and the NUHM (right panels).  Models are divided according to the mass relations as discussed in the text.  
}
\label{fig:poles}
\end{figure}

Returning to the question of the relationship between annihilation mechanism (mass hierarchy) and fine-tuning, 
Figs.~\ref{fig:coann} and \ref{fig:poles} show the $(m_{\neut},\sigma_{SI})$ plane for a variety of subsets of our CMSSM and NUHM parameter spaces chosen by mass relation. Points in Figs.~\ref{fig:coann} and \ref{fig:poles} are color-coded by the value of $\Delta$, the fine-tuning parameter, for each model point.

\paragraph{Stau Coannihilation}

In the top left and right panels of Fig.~\ref{fig:coann} we show the points obeying the mass relation that roughly characterizes morels for which neutralino-stau coannihilation is significant, namely $m_{\neut} \approx m_{\stau}$, in the CMSSM and the NUHM, respectively.  In both cases, the least fine-tuned models are the most accessible to direct detection experiments.  In the CMSSM, there is a clear anti-correlation between $\sigma_{SI}$ and $\Delta$, but in the NUHM the relationship is less clear.  Furthermore, in the CMSSM is appears that all cases with very light $m_{\neut} \approx m_{\stau} \lesssim 180$ GeV would be accessible to XENON-1T\footnote{We note that statements regarding detectability in specific experiments depend on the strangeness content of the nucleon, which is not well-known and can change $\sigma_{SI}$ by a factor of a few~\cite{Ellis:2008hf}. We caution against a strict interpretation of whether a model point is detectable in a particular experiment, but use the sensitivity contours as general guidelines.}, however this conclusion does not hold for the NUHM, where there is considerably more variation in both $\sigma_{SI}$ and $\Delta$.

\paragraph{Stop Coannihilation}

The bottom left and right panels of Fig.~\ref{fig:coann} show the model points obeying the mass relation $m_{\neut} \approx m_{\stop}$, in the CMSSM and the NUHM, respectively.  In both the CMSSM and the NUHM these models are extremely fine-tuned with $\Delta > 1000$.  The large fine-tuning comes from the fact that in order to get $m_{\stop}$ to be low enough to be close to the LSP mass, the running of $m_{\stop}$ from $M_0(M_{GUT})$ must be accelerated.  This can be achieved with a large value of the top trilinear coupling, $|A_t| > 1$ TeV.  These large values of $A_{t}$ also drive $m_{H_{u}}$ to be large and negative. 
One can see from Eq.~\ref{zmass} that in order for EWSB to produce the observed value of $m_Z$, in the CMSSM, a large value of $\mu$ is then required.

Eqs.~\ref{eq:deltapieces} and~\ref{eq:delta} show the strong dependence of $\Delta$ on $\mu$.  To illustrate it even more clearly, consider the case of large $\tan\beta$, when $\Delta = \sqrt{5} \times \mu^2/m_Z^2 + O(1/\tan^2\beta)$.  Clearly, large $\mu$ implies large $\Delta$.  It is obvious then why these models are all very fine-tuned for the CMSSM. The additional freedom in the Higgs sector of the NUHM admits somewhat smaller $\mu$, but overall the fine-tuning is uncomfortably large in both the CMSSM and the NUHM for models with $m_{\neut} \approx m_{\stop}$.

\paragraph{Chargino Coannihilation}

In the top left and right panels of Fig.~\ref{fig:poles} we show the points obeying the mass relation that roughly characterizes neutralino-chargino coannihilation, $m_{\neut} \approx m_{\cha}$, in the CMSSM and the NUHM, respectively.  Since the neutralino LSP is a mixture of bino, and neutral wino and higgsino states while the chargino is a mixture of charged wino and higgsino states, and since gaugino universality implies that that the bino mass is always $\sim 1/2$ the wino mass, in order for $m_{\neut} \approx m_{\cha}$ they must both have significant higgsino components.  When $\mu < M_1$, $m_{\neut} \approx m_{\cha} \approx \mu$, in which case heavier neutralinos and/or charginos will have larger $\mu$ and are therefore more fine-tuned, as is evident in the upper panels of Fig.~\ref{fig:poles} for both the CMSSM and the NUHM.

Since the LSP is a mixed bino-higgsino in these cases, $\sigma_{SI}$ is generally quite large.  There is more variation in $\sigma_{SI}$ in the NUHM than in the CMSSM for the reasons discussed above related to the variation in higgsino content and $m_H$.  In the CMSSM, all models in this category would be detectable by XENON-1T, while this is not quite the case in the NUHM.

\paragraph{Higgs Poles}

Finally, in the bottom panels of Fig.~\ref{fig:poles} we show models in which the LSP is nearly degenerate with the light or pseudoscalar Higgs for the CMSSM (left) and the NUHM (right).  The light Higgs pole occurs where $2 m_{\neut} \approx m_h$, so $m_{\neut} \approx 60$ GeV.  For such a light LSP, it is necessary that both $M_1$ and $\mu$ are small, so the fine-tuning is relatively low in these models.  The heavy Higgs pole occurs where $2 m_{\neut} \approx m_A$.  Again, because of the additional freedom in the Higgs sector in the NUHM, the parameter space for A-pole annihilations is larger than in the CMSSM, resulting in a larger range of $\sigma_{SI}$ in the NUHM than in the CMSSM.  In the CMSSM, A-pole points at lower $m_{\neut}$ and with larger $\sigma_{SI}$, i.e.~the most accessible to direct dark matter searches, are the least fine-tuned. In the NUHM, that conclusion does not hold; points with $\Delta$ as small as a few$\times 10$ have cross sections that may not be accessible even to next generation direct detection experiments.

\begin{figure}[h]
\includegraphics[width=65mm, height=63mm]{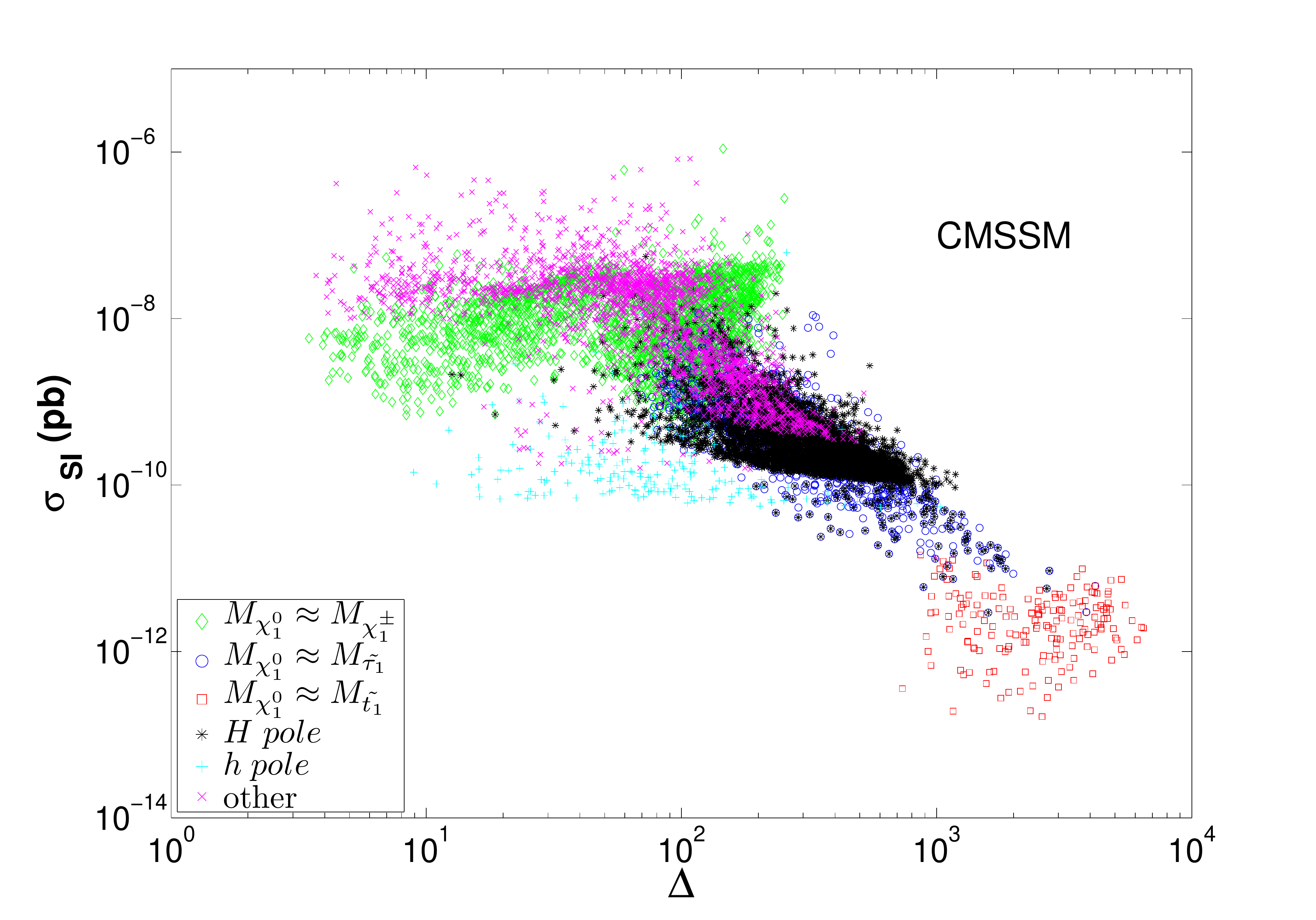}
\includegraphics[width=65mm, height=63mm]{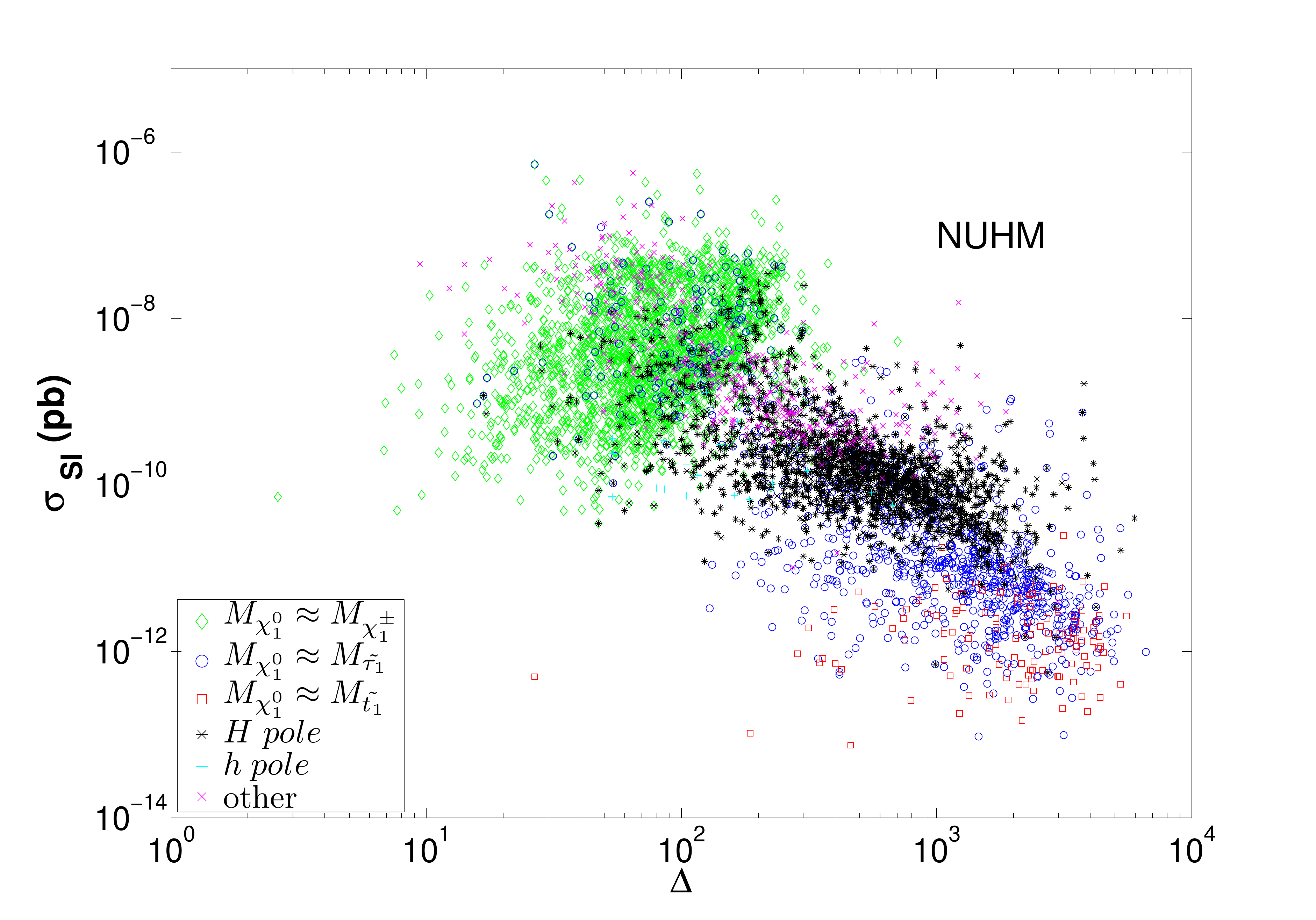}
\caption{Spin-independent neutralino-nucleon scattering cross section, $\sigma_{SI}$, as a function of fine-tuning parameter, $\Delta$, for the CMSSM (left) and
the NUHM (right).  Color-coding indicates the mass relation obeyed as described in the legend.}
\label{fig:bloodybone}
\end{figure}

Finally, the left and right panels of Fig.~\ref{fig:bloodybone} show the spin-indedepent neutralino-nucleon elastic scattering cross section as a function of the fine-tuning parameter in the CMSSM and the NUHM, respectively.   From the general downward slope of the points in the $(\Delta,\sigma_{SI})$ plane, it is evident that as $\Delta$ becomes large, 
$\sigma_{SI}$ tends to decrease in both the CMSSM and the NUHM.
This is related to the fact that large $\Delta$ implies large $\mu$, which, all other factors being fixed, would result in a more bino-like LSP.  Especially in the CMSSM, the least fine-tuned models tend to be the easiest to rule out, with the general trend that increasing sensitivity to $\sigma_{SI}$ will test increasingly fine-tuned models.

In the NUHM, the relation between $\sigma_{SI}$ and fine-tuning does not hold as clearly. For example, models with small fine-tuning and $m_{\neut} \approx m_{\cha}$ may be much more difficult to discover via direct dark matter searches if we are in an NUHM scenario than if the CMSSM is an adequate description of nature. Given the additional freedom in the Higgs sector of the NUHM, it is perhaps surprising that the CMSSM and the NUHM exhibit as many similarities as they do.

%%%%%%%%%%%%%%%%%%%%%%%%%%%%%%%%%%%%%%%%
\section{Implications of Higgs searches and the limit on BR$(B_s \rightarrow \mu^+ \mu^)$}

Thus far we have summarized the results of~\cite{AFS}.  Since the publication of~\cite{AFS}, however, there have been two measurements that have a profound effect on the parameter space of the CMSSM and the NUHM: First, the LHCb Collaboration placed a very strong limit on the branching ratio of $B_s$ of $BR(B_s \rightarrow \mu^+ \mu^-) < 4.5 \times 10^{-9}$ at the 95\% confidence level~\cite{LHCbmm}.  This branching ratio is enhanced at large $\tan\beta$, a general feature of models excluded by the new constraint.  The second important measurement is the discovery of a new particle with properties consistent with those expected of a Standard Model-like Higgs boson with a mass of $\sim 125$ GeV~\cite{higgsCMS,higgsATLAS}.  This relatively large mass, within the CMSSM or NUHM, favors large $A_0$ and large $\tan\beta$~\cite{Ellis:2012aa}. 

\begin{figure}
  \includegraphics[width=.5\textwidth, height=70mm]{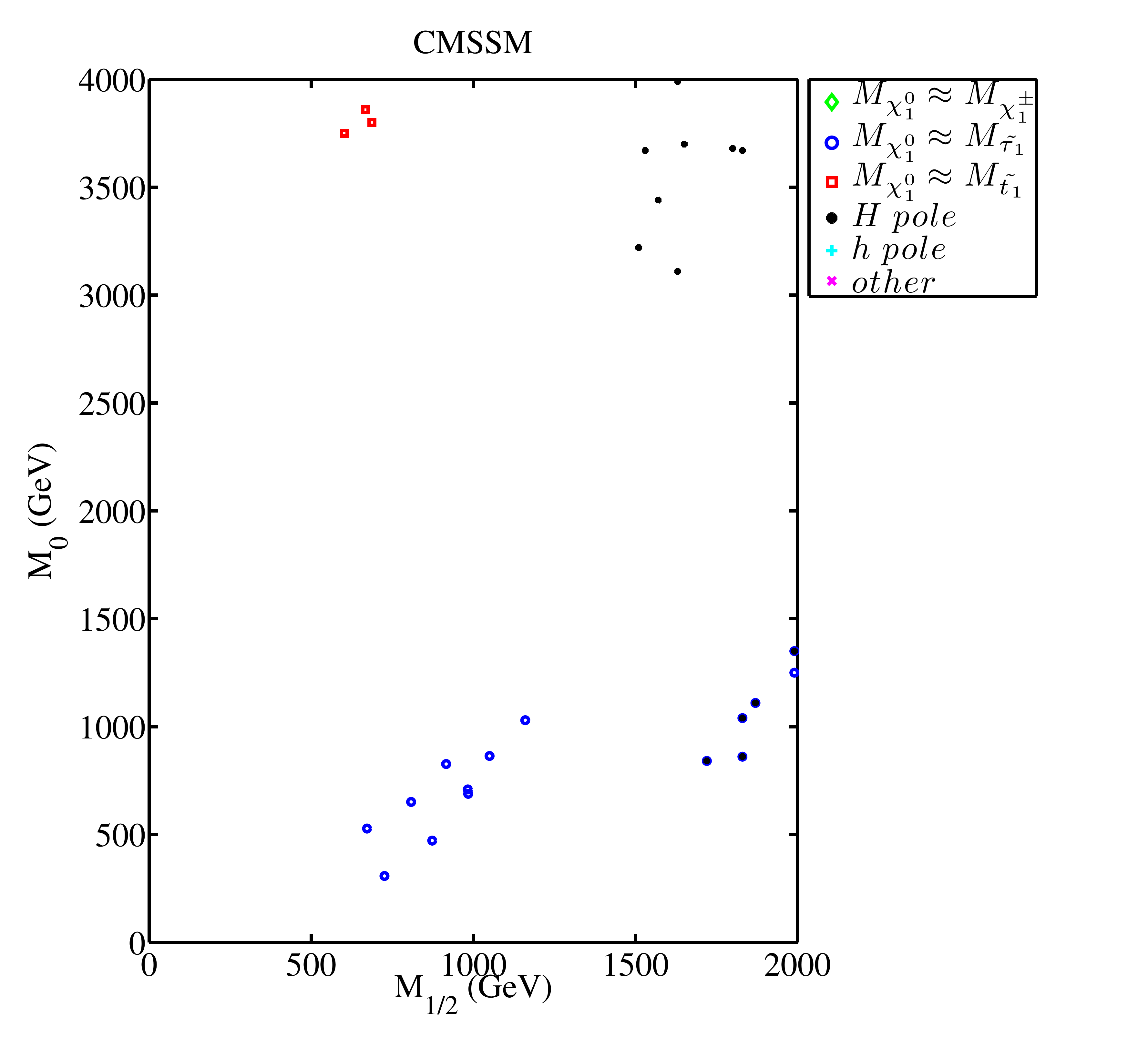}
   \includegraphics[width=.5\textwidth, height=70mm]{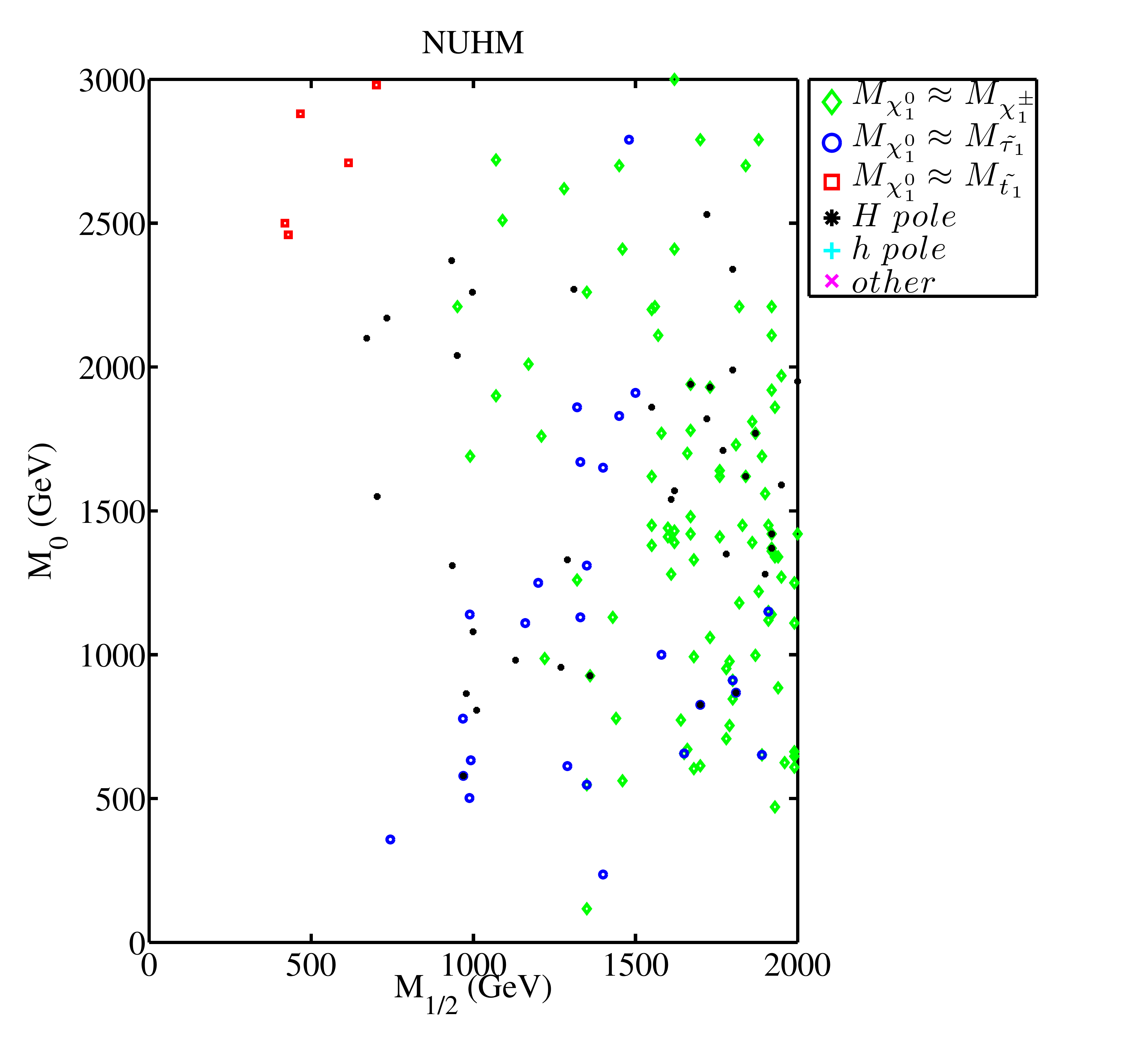}
  \caption{The $(M_{1/2},M_0)$ plane of the CMSSM (left) and the NUHM (right) after applying $BR(B_s \rightarrow \mu^+\mu^-) > 4.5 \times 10^{-9}$ and $123$ GeV $ < m_h < 127$ GeV.  
Models are color-coded by mass relation as described in the legend.}
  \label{fig:m0mhf2}
\end{figure}

To investigate the implications for our results of these recent developments, we demand the $BR(B_s \rightarrow \mu^+\mu^-) < 4.5 \times 10^{-9}$ and $123$ GeV $ < m_h < 127$ GeV.  In Fig.~\ref{fig:m0mhf2}, we show the remaining model points in the CMSSM (left) and the NUHM (right) after implementing these constraints.  Comparison with Fig.~\ref{fig:m0_mhf} shows that, especially in the CMSSM, the combined effect of the limit on the $BR(B_s\rightarrow \mu^+\mu^-)$ and $m_h \approx 125$ GeV leaves few viable model points.  We note that these data were generated prior to indications of a large Standard Model-like Higgs mass, so the parameter ranges were not chosen to optimize the parameter space with $m_h \approx 125$ GeV.  The indication, however, is that the fraction of surviving models is far greater for the NUHM than the CMSSM.  

For example, one can see from Fig.~\ref{fig:m0mhf2} that in the CMSSM no points with a substantial higgsino fraction survive.  The primary cosmologically-viable region of CMSSM parameter space where the neutralino LSP is significantly higgsino-like is known as the focus point region, where $\mu$ is small and $M_0$ is large.  For $M_0 < 4$ TeV as we explore here, we do not find any model points compatible with $m_h \approx 125$ GeV, though the possibility of mixed bino-higgsino dark matter and large enough $m_h$ at larger $M_0$ still exists~\cite{Ellis:2012aa,CMSSMhiggs}.  Among NUHM models, by contrast, there are many surviving model points with $m_{\neut} \approx m_{\cha}$, where the LSP is a mixed bino-higgsino state.

A striking feature in both the CMSSM and NUHM $(M_{1/2},M_0)$ planes is that the light Higgs pole appears to have been excluded.  The constraint on the mass of the Standard Model-like Higgs leads to a dearth of viable models at low $M_{1/2}$ in both the CMSSM and the NUHM.

\begin{figure}
\begin{tabular}{cc}
 \includegraphics[width=65mm, height=63mm]{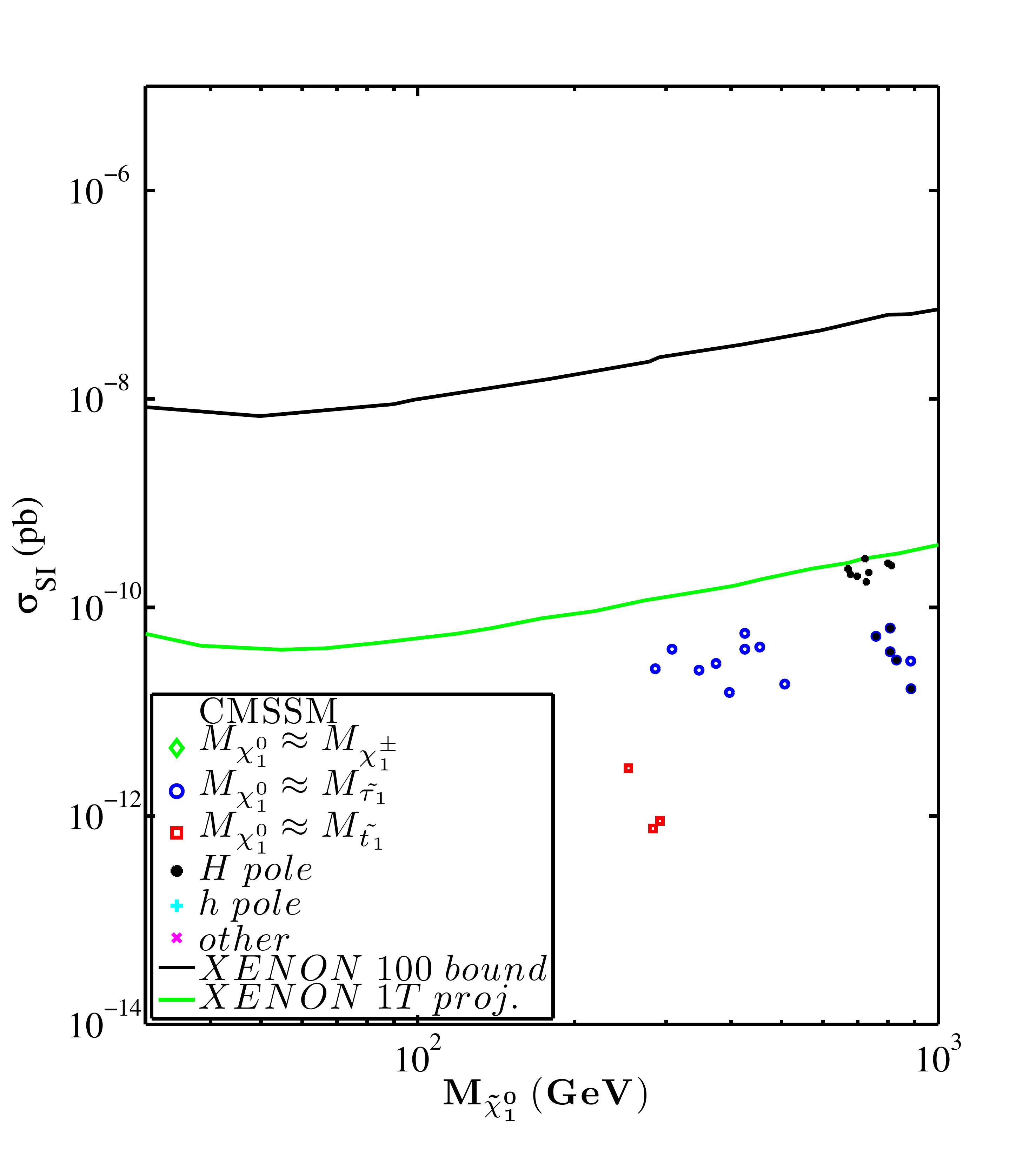} &
   \includegraphics[width=65mm, height=63mm]{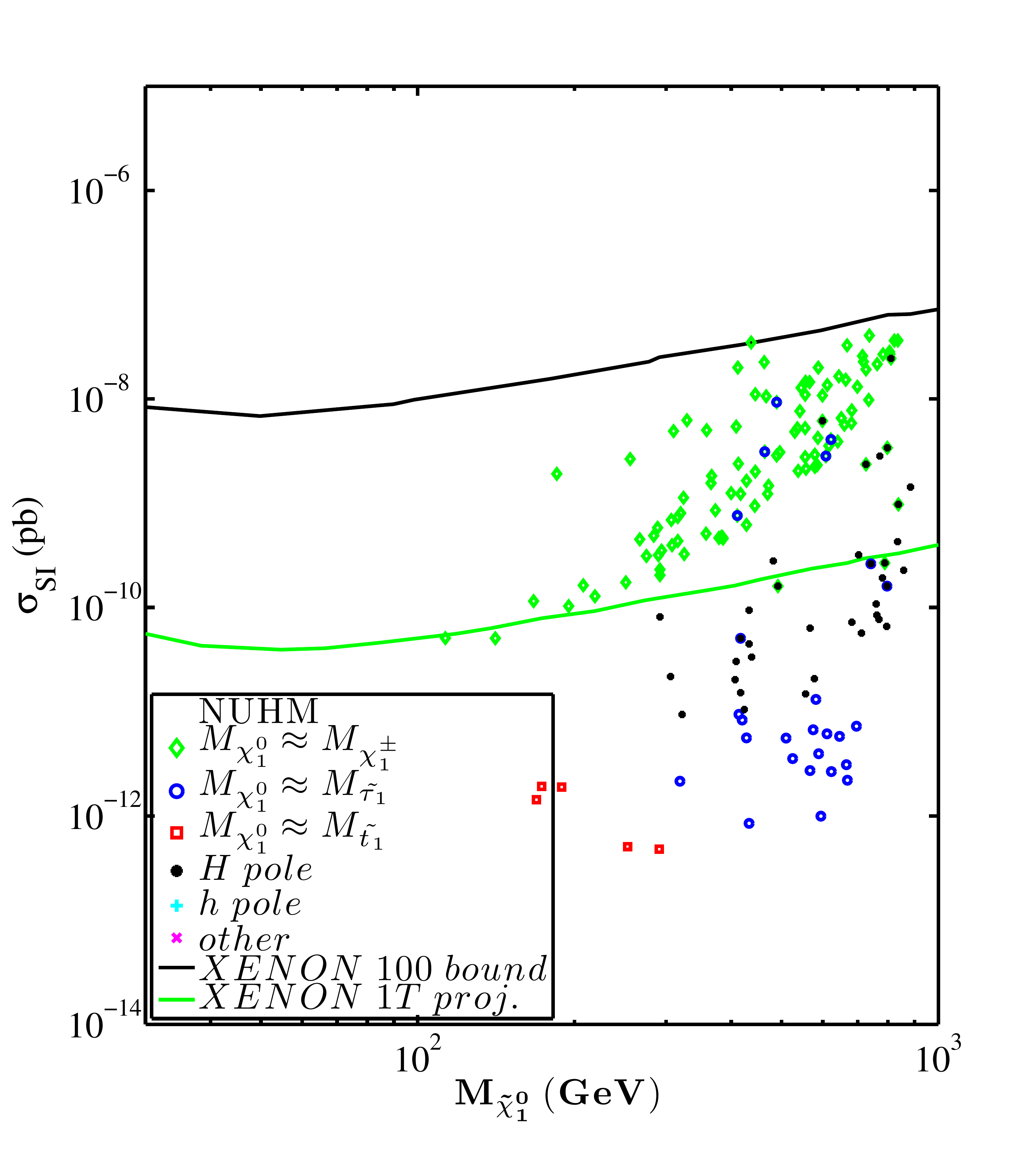} \\
  \includegraphics[width=65mm, height=63mm]{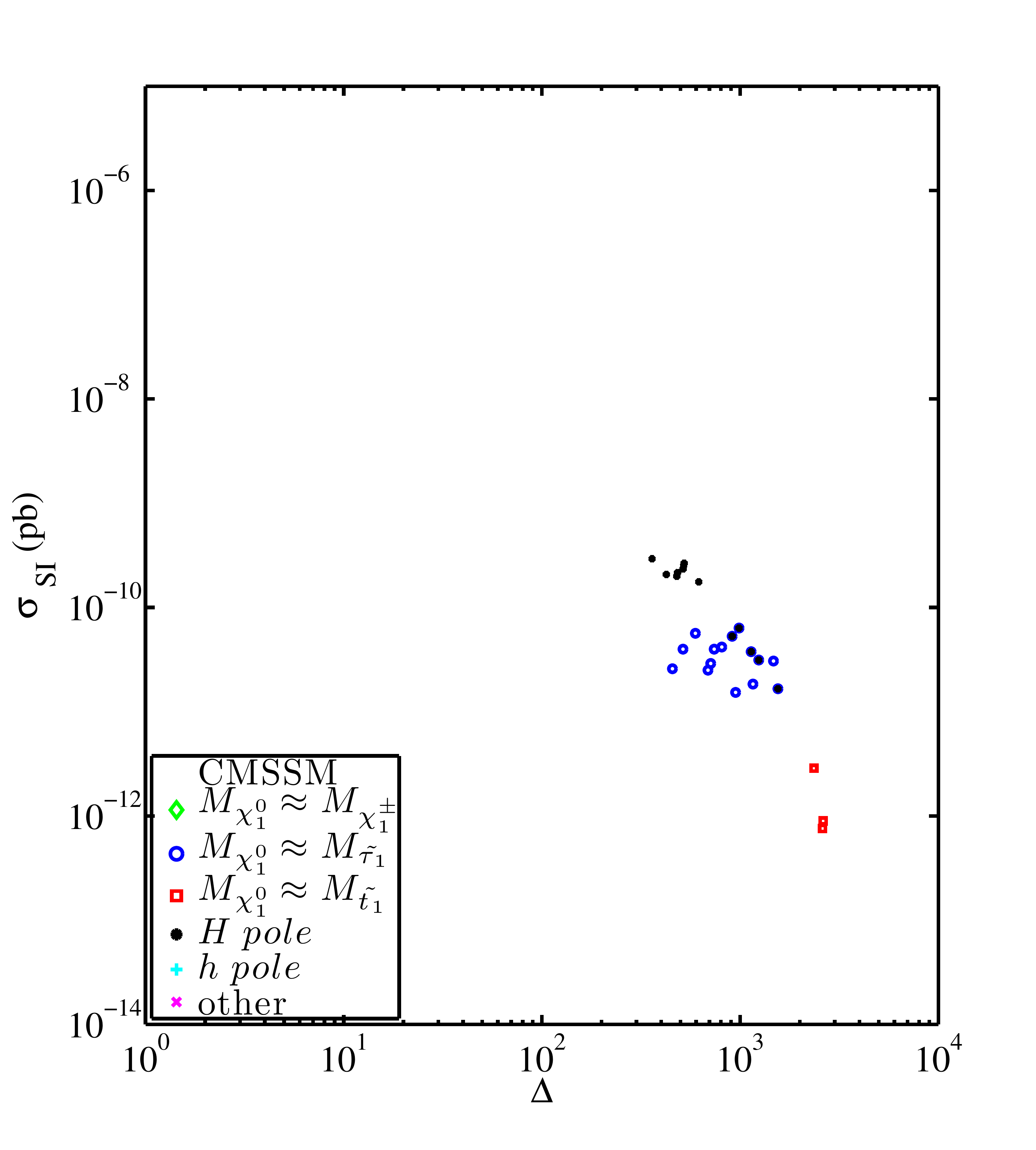} &
   \includegraphics[width=65mm, height=63mm]{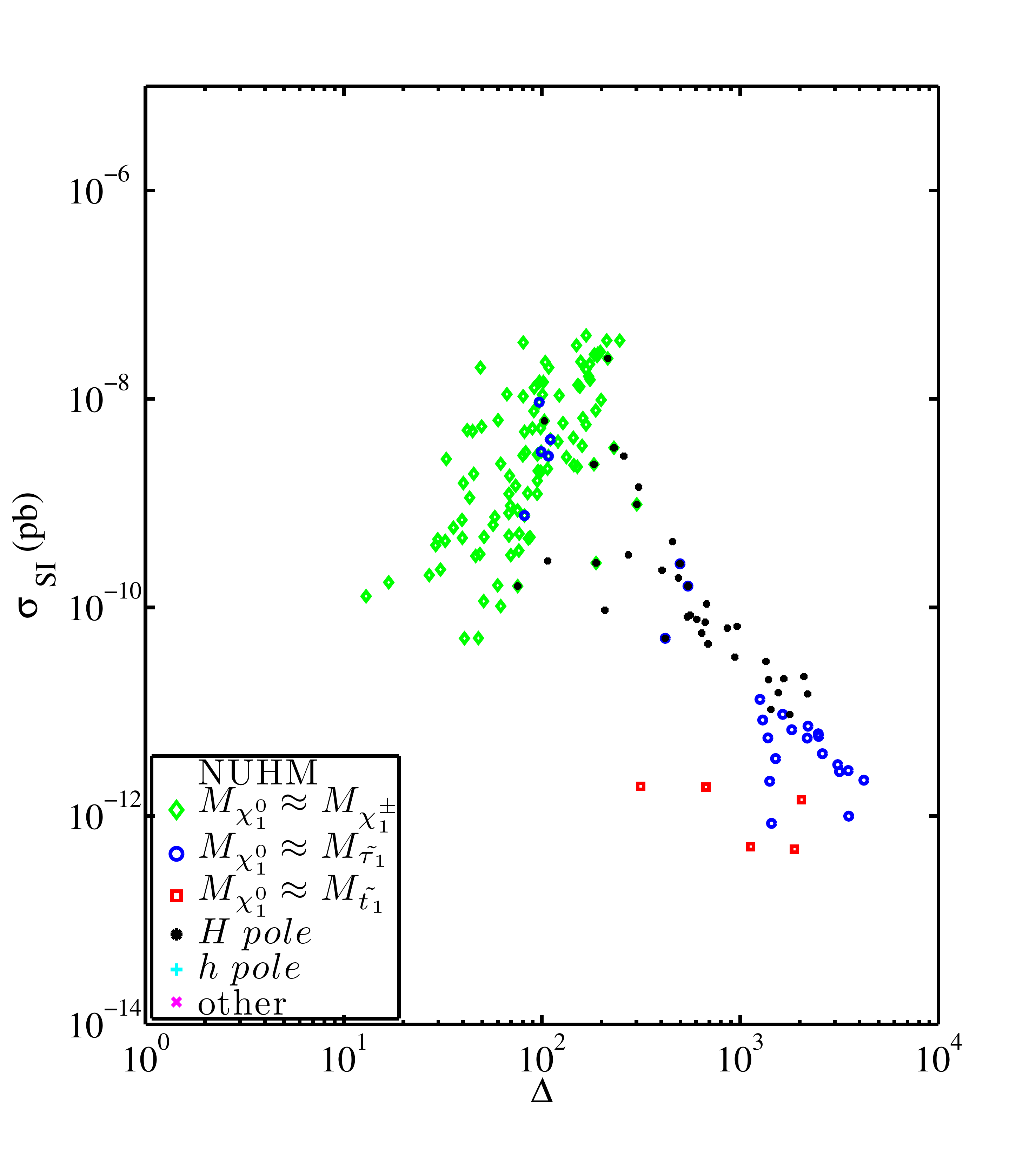}
   \end{tabular}
  \caption{Spin-independent neutralino-nucleon elastic scattering cross section, $\sigma_{SI}$, as a function of neutralino mass, $m_{\neut}$, (top panels) and fine-tuning parameter, $\Delta$, (bottom) for the CMSSM (left panels) and the NUHM (right panels). Model points are color-coded by mass hierarchy as indicated in the legend.}
  \label{fig:higgsbmm}
\end{figure}

Turning to the direct detection prospects, the upper panels of Fig.~\ref{fig:higgsbmm} reveal that in the CMSSM, the models most likely to be discovered by direct dark matter searches are already excluded by the measurement of the Higgs mass and the $BR(B_s \rightarrow \mu^+ \mu^-)$, while in the NUHM, models with $m_{\neut} \approx m_{\cha}$, as well as some scenarios in which $m_{\neut} \approx m_{\stau}$ and/or annihilations are enhanced by the A-pole, may be accessible to XENON-1T or a similar next generation direct dark matter experiment.  Finally, the lower panels of Fig.~\ref{fig:higgsbmm} illustrate that in the CMSSM, the least fine-tuned models have already been excluded; remaining CMSSM points have large fine-tuning and small spin-independent neutralino-nucleon elastic scattering cross sections.  NUHM scenarios, by contrast, still allow for the possibility of low fine-tuning and large enough spin-independent neutralino-nucleon elastic scattering cross sections that there is still reason to be optimistic about the prospects for discovery in next generation direct detection experiments such as XENON-1T.

\section{Summary}

The relationship between the degree of fine-tuning and discoverability of dark matter in current and next generation
direct detection experiments has been investigated.  In~\cite{AFS} it was found that there is considerably more variation in the spin-independent neutralino-nucleon elastic scattering cross section in the NUHM than in the CMSSM, and as a result there is less correlation between the degree of fine-tuning and direct detection prospects in the NUHM than in the CMSSM.  The least fine-tuned CMSSM model points would be the first probed by direct dark matter searches.  In the NUHM, this conclusion is only approximate.  

The relationship between degree of fine-tuning and discoverability of dark matter in direct detection experiments was examined also in light of the specific mechanism(s) by which the relic abundance of neutralino dark matter is suppressed to cosmologically viable values. These mechanisms are modeled by considering mass relations roughly indicative of the regions of parameter space in which a particular mechanism would act to significantly enhance the dark matter annihilation rate in the early universe.  Models with $m_{\neut}\approx m_{\cha}$ may have low fine-tuning and large spin-independent neutralino-nucleon elastic scattering cross section. For $M_{\neut} \sim M_{\stau}$, in the CMSSM most cases with very light $m_{\neut} \approx m_{\stau} \lesssim 200$ GeV would be accessible at XENON-1T or a similar experiment.  For the NUHM, however, it is possible that the lightest neutralino has $\sigma_{SI} \lesssim 10^{-12}$ pb for a large range of $m_{\neut}$.  For the case of $M_{\neut} \sim M_{\stop}$, it is clear that the neutralino dark matter would not be discoverable even with an experiment like XENON-1T, and furthermore almost all of the points in this case are quite fine-tuned with $\Delta > 1000$.

Finally, we investigated the impact on our results of the discovery of a new particle consistent with a Standard Model-like Higgs boson with a mass of $\sim125$ GeV as well as the improvement in the limit on the $BR(B_s \rightarrow \mu^+ \mu^-)$.
We find that these two constraints already exclude the least fine-tuned CMSSM points and that remaining viable parameter space may be difficult to probe with next generation direct dark matter searches.  However, relatively low fine-tuning and good direct detection prospects are still possible in NUHM scenarios.

%%%%%%%%%%%%%%%%%%%%%%%%%%%%%%%%%%%%%%%%%%%%%%%%
%% BACKMATTER
%%%%%%%%%%%%%%%%%%%%%%%%%%%%%%%%%%%%%%%%%%%%%%%%

\begin{theacknowledgments}
P.S. thanks Katherine Freese and Steven Amsel for fruitful collaboration and the Center for Theoretical Underground Physics and Related Areas (CETUP* 2012) in South Dakota for its hospitality and for partial support during the completion of this work.
\end{theacknowledgments}

%%%%%%%%%%%%%%%%%%%%%%%%%%%%%%%%%%%%%%%%%%%%%%%%
%% The bibliography can be prepared using the BibTeX program or
%% manually.
%%
%% The code below assumes that BibTeX is used.  If the bibliography is
%% produced without BibTeX comment out the following lines and see the
%% aipguide.pdf for further information.
%%
%% For your convenience a manually coded example is appended
%% after the \end{document}
%%%%%%%%%%%%%%%%%%%%%%%%%%%%%%%%%%%%%%%%%%%%%%%%

%%%%%%%%%%%%%%%%%%%%%%%%%%%%%%%%%%%%%%%%%%%%%%%%
%% You may have to change the BibTeX style below, depending on your
%% setup or preferences.
%%
%%
%% For The AIP proceedings layouts use either
%%%%%%%%%%%%%%%%%%%%%%%%%%%%%%%%%%%%%%%%%%%%

\bibliographystyle{aipproc}   % if natbib is available
%\bibliographystyle{aipprocl} % if natbib is missing

%%%%%%%%%%%%%%%%%%%%%%%%%%%%%%%%%%%%%%%%%%%
%% The following lines show an example how to produce a bibliography
%% without the help of the BibTeX program. This could be used instead
%% of the above.
%%%%%%%%%%%%%%%%%%%%%%%%%%%%%%%%%%%%%%%%%%%

\end{document}